\documentclass[prb,aps,showpacs,floatfix,twocolumn]{revtex4}

\usepackage{subfigure}
\usepackage{graphicx}
\usepackage{rotating}
\usepackage{amsmath}

\begin{document}

\title{Structural and vibrational properties of two-dimensional $\rm Mn_xO_y$ nanolayers on Pd(100)}

\author{C. Franchini}
\email{cesare.franchini@univie.ac.at}
\affiliation{Faculty of Physics, Universit\"at Wien and Center for Computational Materials Science,  A-1090 Wien, Austria}

\author{R. Podloucky}
\affiliation{Institut f\"ur Physikalische Chemie, Universit\"at Wien and Center for Computational Materials Science, A-1090 Wien, Austria}

\author{F. Allegretti}
\email{francesco.allegretti@uni-graz.at}
\affiliation{Institute of Physics, Surface and Interface Physics, Karl-Franzens University Graz,
A-8010 Graz, Austria}

\author{F.Li, G. Parteder, S. Surnev, F.P. Netzer}
\affiliation{Institute of Physics, Surface and Interface Physics, Karl-Franzens University Graz,
A-8010 Graz, Austria}


\begin{abstract}

Using different experimental techniques combined with density functional based theoretical methods we have explored
the formation of interface-stabilized manganese oxide structures grown on Pd(100) at (sub)monolayer coverage.
Amongst the multitude of phases experimentally observed we focus our attention on four structures which can be
classified into two distinct regimes, characterized by different building blocks.
Two oxygen-rich phases are described in terms of MnO(111)-like O-Mn-O trilayers, whereas the other two have a
lower oxygen content and are based on a MnO(100)-like monolayer structure.
The excellent agreement between calculated and experimental scanning tunneling microscopy images and vibrational 
electron energy loss spectra allows for a detailed atomic description of the explored models.

\end{abstract}
\maketitle

\section{Introduction}

Ultra-thin oxide films grown on metal surfaces with thickness in the nanometer range are promising systems, where 
the reduced dimensionality and the dominance of surface and interface effects may lead to novel physical and chemical 
properties, intrinsically different from those of the corresponding bulk counterparts. As such, low-dimensional oxide 
films with tailored properties are increasingly exploited in technological applications, as for example in micro- and 
nano-electronics, magnetic recording and heterogeneous catalysis.

From a structural point of view, systems only a few atomic layers thick are often different from the bulk-like phases 
in terms of local coordination, long-range ordering and surface symmetry. As a consequence of the interaction with the 
support, even terminations which are not favored as bulk-truncated surfaces may be stabilized, and depending on the 
growth conditions several different structures appear.

For 3$d$ transition metal oxides, numerous structural studies of interfacial phases on various substrates are available 
in the literature, such as FeO on Pt(111) \cite{ritter, giordano}, Pt(100) \cite{shaik} and Ru(0001) \cite{kett}, 
vanadia on Pd(111) \cite{surnev1, surnev2, surnev3} and Rh(111) \cite{hannes1, hannes2, hannes3}, Ni oxide on 
Pd(100) \cite{nioleed, nioexp} and Ag(100) \cite{atr1, atr2}, and Mn oxide on Pt(111) \cite{widdra}. In the present study we 
focus on the Mn oxide phases formed at a Mn coverage of up to one monolayer (ML) on Pd(100).  We have previously characterized the 
high thickness regime (15-20 ML), showing that MnO(100) with bulk-like in-plane lattice constant is stable in a wide range of pressure 
and temperature \cite{1001}; however, at more reducing conditions MnO with (111) orientation can be grown. The MnO(111) surface 
is polar, and at high thickness the films tend to become unstable towards faceting: as a result, the morphology is dominated by 
triangular pyramids exposing the lowest-energy (100) facets \cite{1001}. At the other extreme of oxidizing conditions, we 
have demonstrated that flat MnO(100) films can be preferentially oxidized to yield $\rm Mn_3O_4$(001) \cite{noimn3o4}. 

\begin{figure*}[htb]
\centering
\includegraphics[clip=,width=1.0\linewidth]{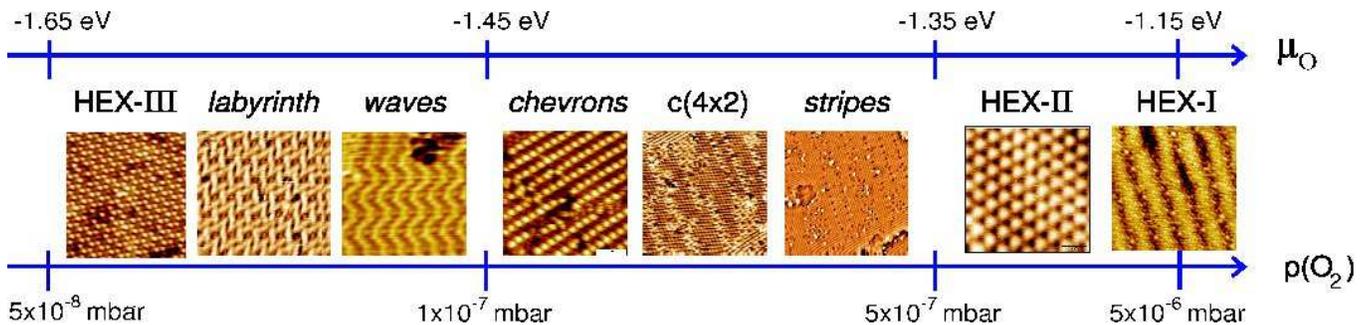}
\caption{Schematic phase diagram of the interfacial Mn oxides, presented as a function of the oxygen pressure P(O$_2$) and 
of the oxygen chemical potential $\mu_O$ \cite{stmli}. The nominal coverage of Mn on Pd(100) is $\sim$ 0.75 ML.}
\label{fig:1}
\end{figure*}

Below 1 ML, a complex phase diagram of Mn oxides on Pd(100) is observed \cite{stmli}, composed of nine different interface-stabilized phases. 
The phase diagram as a function of the oxygen pressure P($\rm O_2$) and of the O chemical potential $\mu_O$,
is depicted schematically in Fig. \ref{fig:1}. It comprises: 

- two hexagonal phases (HEX-I and HEX-II), which are both obtained at high $\rm O_2$ pressures (above 5$\times10^{-7}$ mbar) 

-  a c(4$\times$2) structure and a {\em stripe} phase described as a uni-axially compressed c(4$\times$2), which are both 
   stabilized at intermediate $\rm O_2$ pressures (around 5$\times10^{-7}$ mbar)

- two structures which were called {\em chevrons} (CHEV-I and CHEV-II), because of their STM appearance

- two reduced phases with complex structures, named {\em waves} and {\em labyrinth}

- finally, at the most reducing conditions, a third hexagonal phase (HEX-III), commensurate with the Pd(100) substrate ($\times$2) 
  along one of the two $<011>$ directions. 

Here, we aim to rationalize this extraordinary architectural flexibility of the interfacial Mn oxides on Pd(100) in terms of 
realistic structural models. On the basis of high-resolution electron energy loss spectroscopy (HREELS) data we show that it is possible 
to distinguish two different oxygen regimes: an O-rich regime, comprising the HEX-I and HEX-II phases
and characterized by a dominant phonon loss at $\approx$ 70 meV, and an O-poor regime encompassing all other phases, which display a 
single phonon loss at $\approx$ 44 meV. Theoretical models based on an educated guess of possible structures are tested and directly 
compared with scanning tunneling microscopy (STM), low energy electron diffraction (LEED) and HREELS results. In the following, we 
consider in detail four phases, c(4$\times$2), CHEV-I, HEX-I and HEX-II, for which the atomic scale modeling represents a 
computationally affordable task due to the relatively small size of the surface unit cell. Through the combined theoretical and 
experimental investigation, it is concluded that the two O-rich phases can be described in terms of O-Mn-O trilayers with 
a MnO(111)-like structure, whereas the O-poor regime is based on a compressed MnO(100)-like model. We find that the high degree of 
flexibility of these two basic models is favored by the formation of regular arrays of vacancies, in line with previous investigations
on related systems \cite{nioexp, widdra}. Importantly, our description provides a natural link between the interfacial phases and the 
MnO(100) and MnO(111) films of the high thickness regime.

The paper is organized as follows: the experimental procedures and computational details are described in Section
\ref{one1} and \ref{one2}, respectively. In Section \ref{two1} we discuss the main experimental evidence for the existence of two 
different oxygen pressure regimes; full structural models for the c(4$\times$2) and CHEV-I of the O-poor or intermediate-O regime are 
presented in Section \ref{two2}, while the proposed atomic structure for the HEX-I and HEX-II phases of the O-rich regime is discussed 
in section \ref{two3}. Finally, section \ref{three} is devoted to the summary and conclusions. 

\section{Technical Aspects}
\label{one}

\subsection{Experimental Procedure}
\label{one1}
The experiments have been conducted in a number of different ultrahigh vacuum (UHV) chambers with typical base pressure 
p = 1-2$\times10^{-10}$ mbar.  The structure and morphology  of the Mn oxides films were characterized by STM measurements 
performed in our home-laboratories in Graz, in two different custom-designed systems. One STM system comprises a 
variable-temperature STM (Oxford Instruments), a LEED optics, a  cylindrical mirror analyzer for Auger electron spectroscopy (AES), 
and the usual facilities for crystal cleaning and physical vapor deposition. The other STM system is equipped with a variable-temperature 
AFM-STM Omicron apparatus and a LEED optics. All STM images were recorded in constant current mode at room temperature (300 K) with 
electrochemically etched W tips, which were cleaned {\em in situ} by electron bombardment. These measurements were complemented by 
spot-profile analysis low energy electron diffraction (SPA-LEED) experiments performed in an independent UHV chamber, which has been 
described in more detail elsewhere \cite{spaleed}. HREELS measurements were performed to elucidate the phonon structure of the oxide layers, 
using an ErEELS spectrometer \cite{hreels},  which was operated  in specular reflection geometry at 60\ensuremath{^\circ} incidence from 
the surface normal and with a primary energy of  6.5 eV.  The crystal was kept at RT during the measurements, and a typical resolution of 6 meV 
was ensured, as measured at the full width at half maximum of the reflected primary peak. LEED provided a common link with the structures 
prepared in the various chambers.

The clean Pd(100) crystal surface was prepared  by the usual combination of 1 keV Ar$^+$-ion sputtering and annealing at 1000 K. Further 
annealing for 3-5 min. in $\rm O_2$ atmosphere (2$\times$10$^{-7}$ mbar) at 570 K, followed by a brief flash to 1000 K in UHV, was used to remove 
contamination of residual atomic carbon. The preparation protocol for the interfacial manganese oxide layers on Pd(100) consists of deposition 
of Mn metal in UHV with the sample kept at room temperature (RT), followed by post-oxidation (PO) at higher temperature in a molecular oxygen 
atmosphere. This PO procedure has been preferred to the reactive evaporation (RE) procedure involving deposition of Mn in oxygen atmosphere, 
because it typically provides films with smoother morphology and containing a single dominant phase, as judged by STM and LEED \cite{stmli}. 
The Mn deposition rate was monitored by a quartz crystal microbalance, and evaporation rates of 0.5 ML min$^{-1}$ or less were usually employed,  
1 ML containing 1.32$\times$10$^{15}$ Mn atoms/cm$^2$  as referred to the Pd(100) surface atom density.  $\rm O_2$ pressures in the range  
1$\times$10$^{-7}$-5$\times$10$^{-6}$ mbar \cite{mark1} and sample temperatures in the range 570-770 K were used, the exact combination of 
these parameters crucially depending on the desired long-range-ordered phase.  

Of the four different phases considered in detail in this work, the HEX-I phase is obtained by post-oxidation of 0.75 ML Mn at 670 K at the
highest oxygen pressure employed (5$\times$10$^{-6}$ mbar $\rm O_2$), while the CHEV-I structure is typically prepared by post-oxidizing the same 
amount of Mn at 700 K and low $\rm O_2$ pressure (1$\times$10$^{-7}$ mbar). Transition between these two phases, and also to the other 
intermediate phases c(4$\times$2) and HEX-II, is made possible by annealing the HEX I phase in UHV or low $\rm O_2$ pressure at 650-700 K, 
or by performing a second post-oxidation of the CHEV-I phase at increasing $\rm O_2$ pressure.  

\subsection{Computational Details}
\label{one2}
The present calculations were carried out using the projector-augmented-wave (PAW)\cite{paw1,paw2} based
Vienna {\em Ab initio} Simulation Package (VASP)\cite{vasp} in the framework of standard (Kohn Sham theory) 
and generalized (hybrid) density functional theory (DFT). The choice of adopting a method beyond standard DFT 
is due to the complex nature of the system to be described. Although the metallic character of the Pd support is correctly 
described within DFT, this is not true for the correlated nature of the magnetic $\rm Mn_xO_y$ overlayers. 
In a series of recent papers\cite{1001, mnxoy, mno, 1002} we have provided a detailed description
of the obstacles that standard DFT faces when applied to manganese oxides. We have shown that the 
application of methods beyond DFT such as HSE (Heyd-Scuseria-Ernzerhof) is essential to correct the self-energy related DFT  
drawbacks and to provide a satisfactory description of the basic structural, electronic, vibrational 
and magnetic properties of these systems.    

Standard and hybrid DFT differ in the construction of the many body exchange and correlation functional.
In the Kohn Sham approach we have adopted the generalized gradient approximation scheme according to
Perdew, Burke and Ernzerhof (PBE)\cite{pbe}, whereas in hybrid DFT (DFTh) a suitable mixing of Hartree-Fock (HF) 
exchange with the exchange expression of PBE has been used following the HSE03 scheme\cite{hse,hsevasp}. 
The resulting expression for the exchange (x) and correlation (c) energy is
\begin{equation}
E_{xc}^{\rm HSE03} = \frac{1}{4}E_{x}^{\rm HF,sr,\mu} +
        \frac{3}{4}E_{x}^{\rm PBE,sr,\mu} + E_{x}^{\rm PBE,lr,\mu} + E_{c}^{\rm PBE},
\end{equation} in which (sr) and (lr) denote the short- and long-range parts of the
respective exchange interactions. The parameter $\mu$ controls the range
separation: the optimum distance beyond which the short-range interactions
become negligible corresponds to values of $\mu$ between 0.2 and 0.3
$\rm \AA^{-1}$. We have used $\mu=0.3$ $\rm \AA^{-1}$. 

The Pd(100) substrate was modeled by repeated asymmetric four layers thick slabs 
which were used as support for the adsorbed $\rm Mn_xO_y$ phases. 
Most of the calculations have been performed by using 2D (4x2) and (5x3) unit cells
employing the theoretically optimized 2D lattice constant. The corresponding Brillouin 
zone integration was done using 6$\times$6$\times$1 and 6$\times$4$\times$1 $\vec{k}$-point grids, respectively, 
following the Monkhorst-Pack generation scheme. 
Relaxation of all atomic positions were allowed by minimizing the atomic forces but fixing the 
position of atoms in the two deepest Pd layers. 
Albeit the magnetic ordering of the Pd(100)-supported manganese oxide overlayers is unknown,
throughout our theoretical investigation we have included spin polarization which allows for a more
accurate description of these systems.
However, to reduce the already high computational costs, 
we have limited our inspection to the ferromagnetic (FM) alignment, unless otherwise stated. 
The vibrational properties have been calculated by diagonalizing the dynamical matrix of the $\rm Mn_xO_y$
sub-systems. STM images have been simulated by determining iso-surfaces of constant charge, thus following
loosely the Tersoff and Hamann method \cite{stm}.

Finally, the stability of the most relevant models was underpinned by evaluating the change of the
generalized adsorption energy $\gamma$ upon different concentration of manganese and oxygen atoms, namely:
\begin{equation}
\gamma = (E_{slab} - E_{Pd(100)} - n_{Mn}\mu_{Mn} -n_O\mu_O)
\label{ce}
\end{equation}
$E_{slab}$ and $E_{Pd(100)}$ are the energies of the explored model and of the clean Pd(100) slab, respectively, whereas
$n_{Mn}$ and $n_O$ are the number of Mn and O atoms for a given model.
The reference Mn energy, $\mu_{Mn}$ is set to the energy of fcc bulk Mn atoms ($\gamma$-Mn), while $\mu_{O}$
is treated as an extrinsic thermodynamic variable which can be related approximately to the experimental
conditions under consideration ($\mu_{O}$ = -1.2 eV, with respect to the free $\rm O_2$ dimer). We have used a
very similar approach for studying the surface phase diagram of MnO(111)\cite{111}. A detailed description
on the thermodynamic formalism has been presented by Reuter and Scheffler\cite{reuter}.

The HEX-I and HEX-II phases experimentally observed, are found to be incommensurate with (or to have a large coincidence mesh with respect to) 
the Pd(100) substrate. Therefore, specific to this case, we have performed free-standing calculations (i.e. unsupported $\rm Mn_xO_y$ 
thin layers). This is a reasonable assumption because the interaction of oxide overlayers
with the metallic substrate is found to be relatively weak compared to the interactions within the oxide itself\cite{vo}.
For this case the stability of a particular structure was evaluated by calculating the formation energy:
\begin{equation}
E_f = (E_{slab} - n_{Mn}\mu_{Mn} -n_O\mu_O)/(n_{Mn} + n_O)
\label{fe}
\end{equation}

\section{Results and Discussion}
\label{two}

\subsection{Evidence for two sub-monolayer regimes}
\label{two1}

\begin{figure}
\includegraphics[clip=,width=1.0\linewidth]{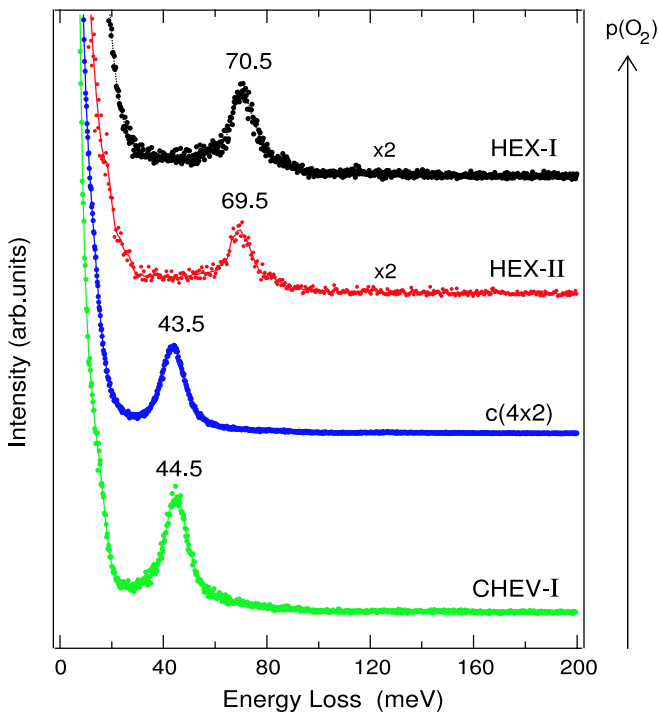}
\caption{HREELS phonon spectra of the four Mn oxide submonolayer phases discussed  in this paper. From the top to the bottom: HEX-I, HEX-II, 
c(4$\times$2) and CHEV-I.}
\label{fig:2}
\end{figure}

As we have briefly mentioned in the introduction, the phase diagram of the interfacial Mn oxides on Pd(100) is rather 
complex and made up of several distinct phases, which have at first sight very little in common as far as their long-range ordering is concerned. 
Identifying common architectural aspects in these self-organized oxide systems is a major step on the way to rationalize the noticeable complexity 
and to unravel the driving factors of the architectural flexibility. HREELS offers a valuable diagnostic tool, because it provides the phonon 
loss spectrum, which is tightly related to the local arrangement and ion coordination in the smaller building blocks constituting each long-range 
ordered phase\cite{hreels-units}. In particular, common building blocks shared between different oxide phases are expected to 
result in similar phonon losses, rendering HREELS sensitive to the occurrence of common structural features. 

\begin{figure}[htb]
\includegraphics[clip=,width=1.0\linewidth]{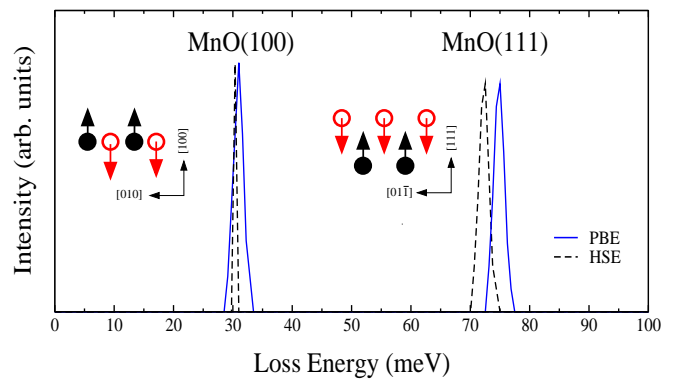}
\caption{Calculated loss peak spectra for the MnO(100) and MnO(111) unsupported thin oxide layers described 
in the text. The circle sketches provide a simplified view of the eigenvectors connected to the dipole active modes.
Filled and empty circles refer to Mn and O atoms, respectively.}
\label{fig:3}
\end{figure}

The HREELS spectra for the four phases considered in the present work are reported in Fig. \ref{fig:2} in the energy range up to 200 meV.
All spectra exhibit a clear single peak structure. The phonon loss is centered at around 70 meV for the two hexagonal phases and shifts down 
to 44-45 meV for the c(4$\times$2) and CHEV-I phases. Interestingly, the phonon spectra of the CHEV-II, {\em waves} and HEX III phases (not shown), 
obtained by further lowering the O chemical potential, are also characterized by a single peak at 43-45 meV. These findings provide clear indication 
of two distinct regimes, an O-rich regime comprising phases with a higher energy phonon loss (70 meV) and an O-poor regime with a single phonon loss 
around 44 meV. Within each regime, we may surmise that phases with different long-range order possess similar building blocks\cite{lund}.

\begin{figure*}[htb]
\begin{minipage}[b]{0.49\linewidth}
\centering
\includegraphics[clip=,width=1.0\linewidth]{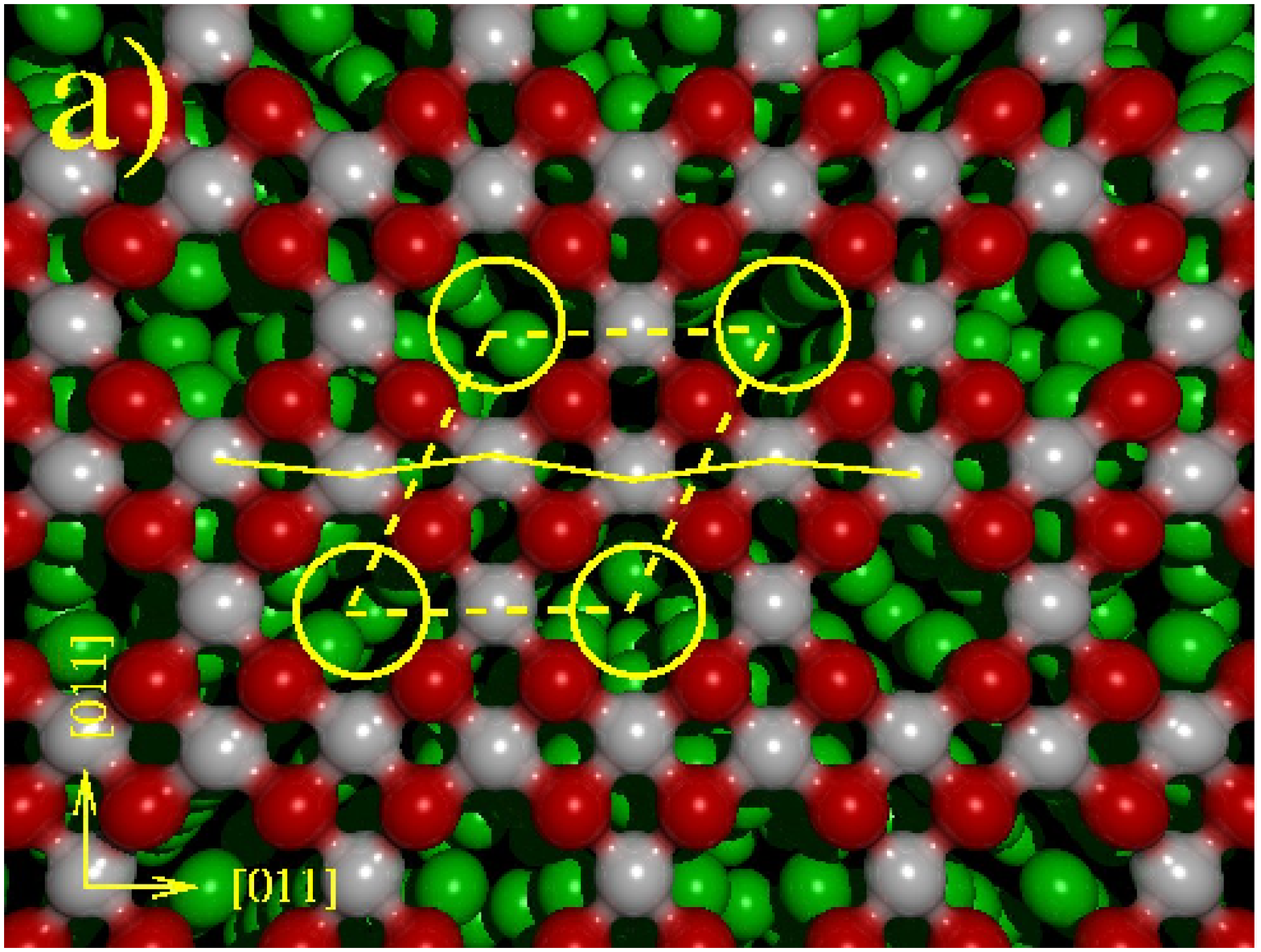}
\end{minipage}
\begin{minipage}[b]{0.49\linewidth}
\centering
\includegraphics[clip=,width=0.75\linewidth]{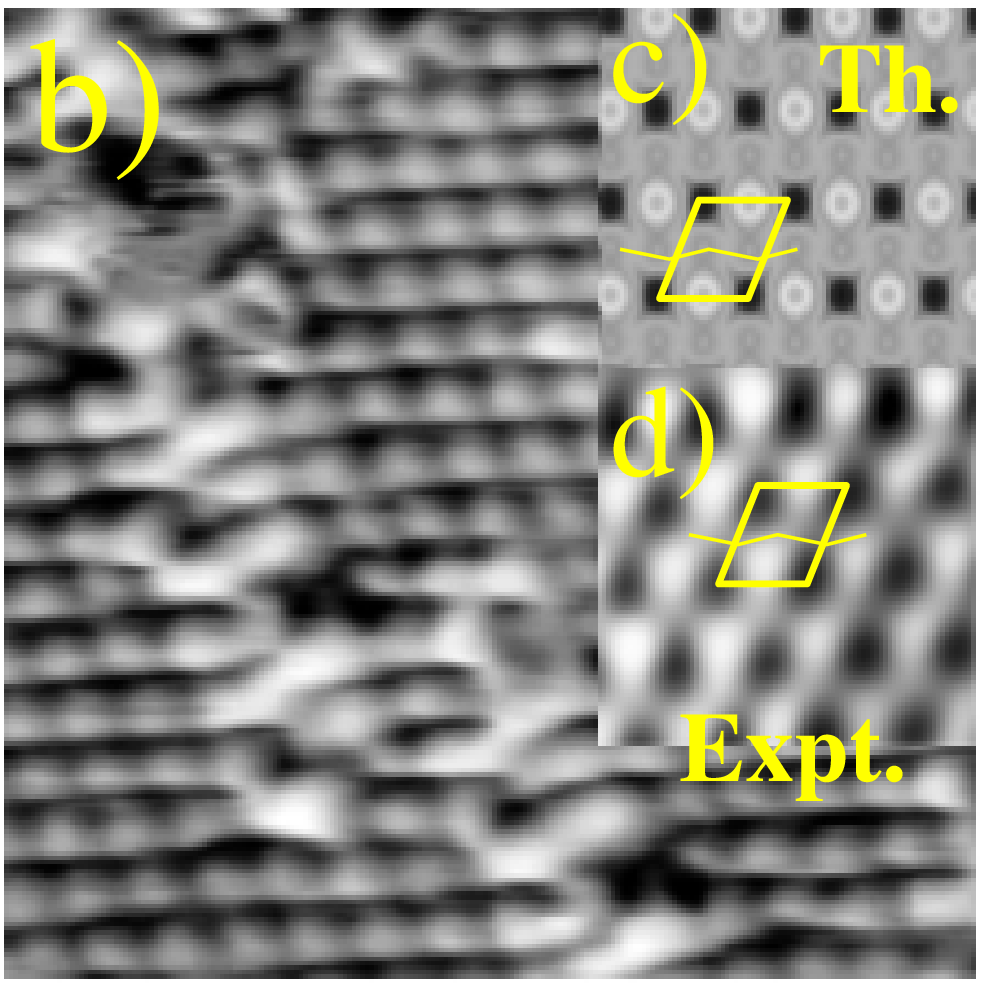} 
\end{minipage}
\begin{minipage}[b]{0.98\linewidth}
\centering
\vspace{1.0cm}
\includegraphics[clip=,width=0.6\linewidth]{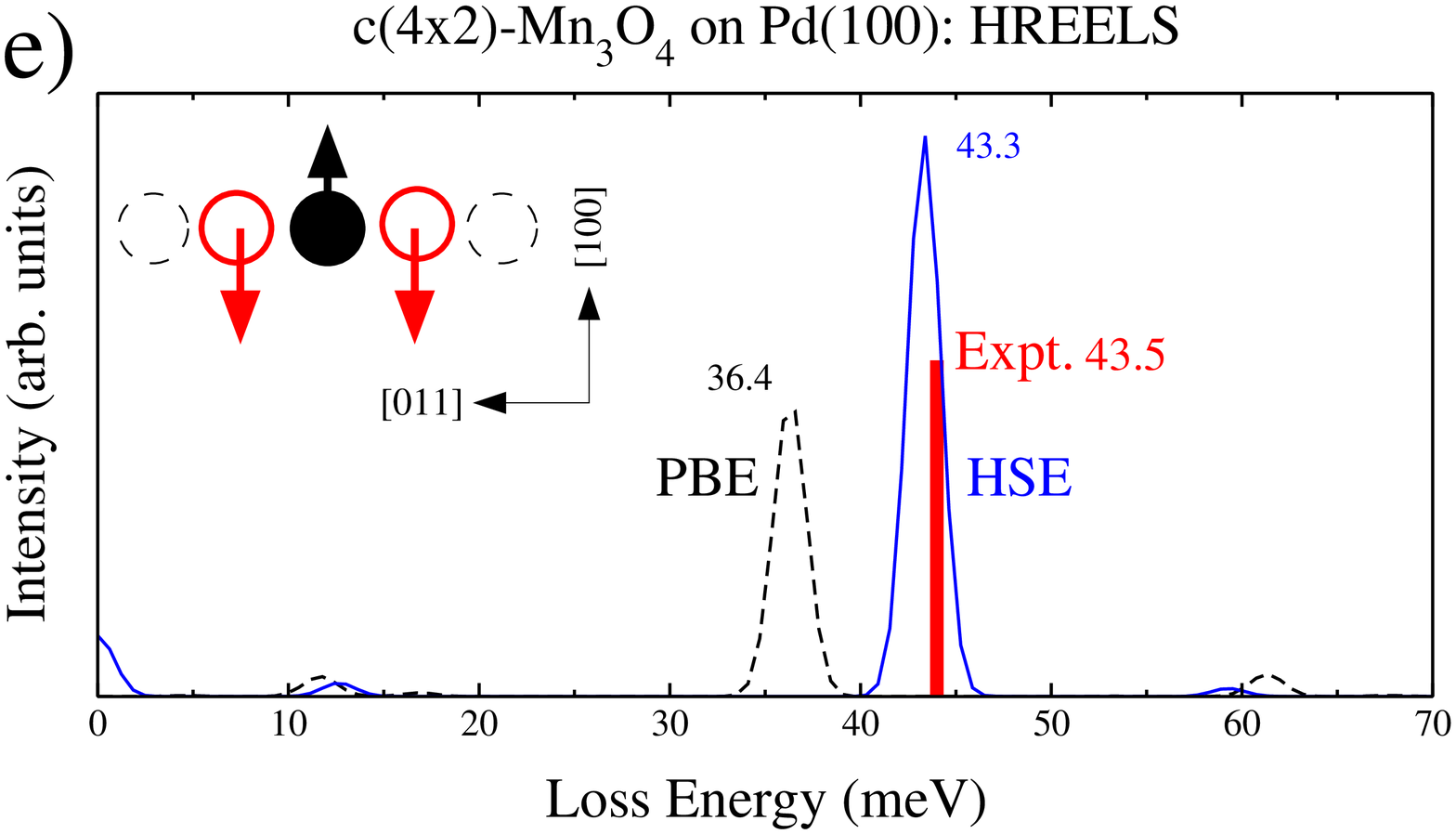}
\end{minipage}
\caption{
(a) Top view of the geometrical model for the Pd(100)-supported c(4$\times$2)-$\rm Mn_3O_4$ phase: 
red (dark gray) spheres: O atoms;  light gray spheres: Mn atoms; green (gray) small spheres: Pd atoms underneath.
Dashed lines delimit the 2D unit cell, the full line indicates the manganese lateral displacements and circles 
highlight the position of the vacancies. 
(b) Experimental STM image of the c(4$\times$2)-$\rm Mn_3O_4$ phase; 
sample bias U = +0.5 V; tunneling current I = 0.2 nA. 
(c-d) High resolution simulated (c) and experimental (d) STM images taken at sample bias U = +1.0 V; tunneling current I =0.25 nA. The theoretical 
image has been calculated considering tunneling into empty states between 0 and +0.5 eV. Full lines indicate the 2D unit cell and the zigzag Mn chain.
(e) Comparison between the measured HREELS phonon value (vertical bar) and PBE (dashed line) and HSE (full line) predicted dipole active modes for 
the  c(4$\times$2)-$\rm Mn_3O_4$ phase. The inset schematically depicts the atomic displacements giving rise to the calculated phonon peaks.
Mn and O atoms are sketched as filled and empty circles, respectively. The dashed line circles indicate the vacancies.}
\label{fig:4}
\end{figure*}

\begin{figure*}[htb]
\centering
\includegraphics[clip=,width=1.0\linewidth]{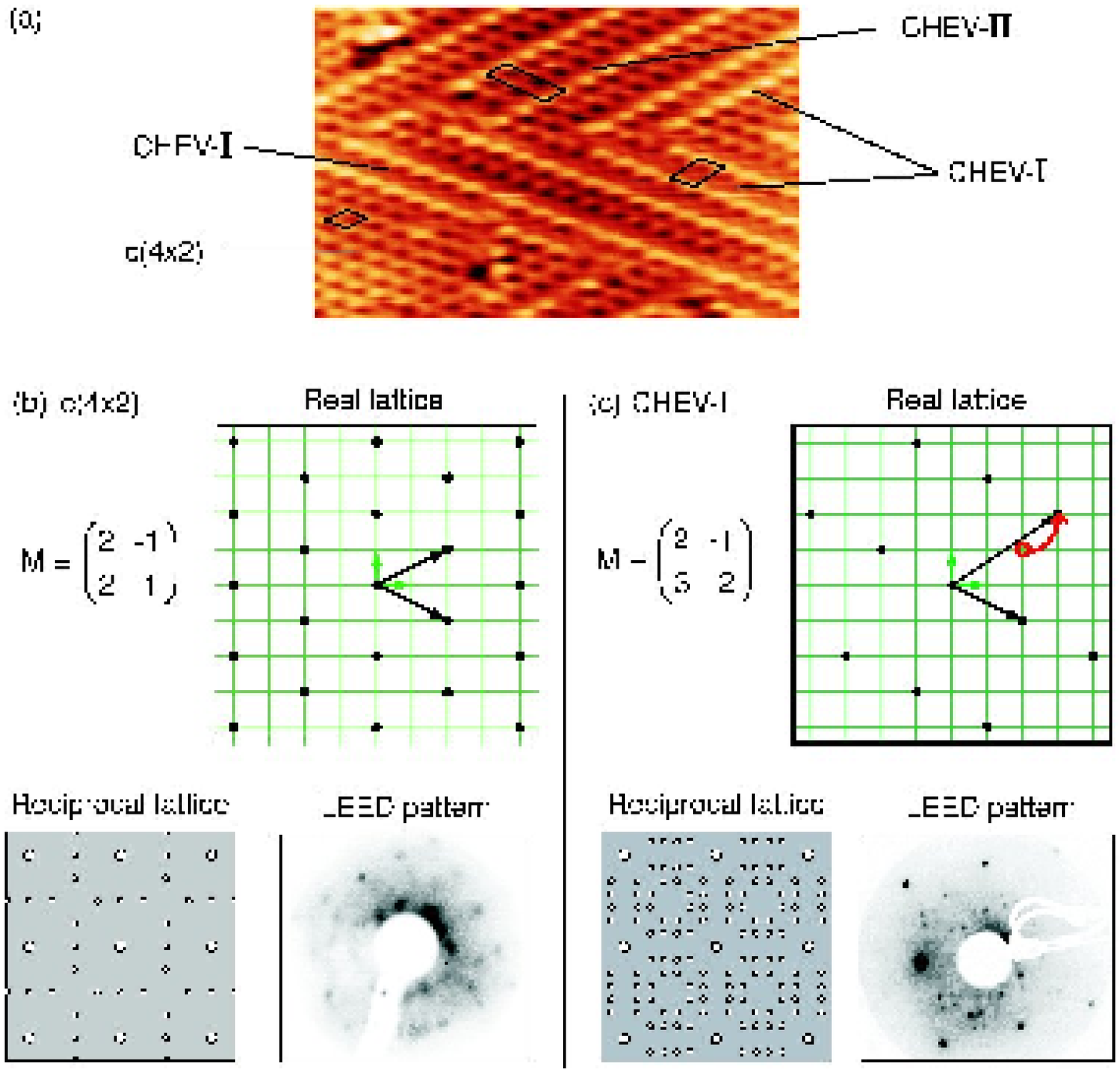}
\caption{(a) STM image showing the characteristic appearance of the {\em chevrons} structures and the local coexistence of the c(4$\times$2), CHEV-I and 
CHEV-II phases (140 \AA $\times$100 \AA ; U = +1.0 V; I = 0.2 nA). (b,c) Real lattice, reciprocal lattice and experimental LEED pattern (E=110 eV) 
for the c(4$\times$2) (panel (b)) and CHEV-I (panel (c)) structures. The red arrow in the right top panel shows how the periodicity of the CHEV-I 
structure can be obtained from the c(4$\times$2) periodicity by extending one of the two lattice vectors to a neighboring anti-phase position.}
\label{fig:5}
\end{figure*}

Based on the hexagonal symmetry of the more oxidized (HEX-I and HEX-II) phases and on the lattice constant of the HEX-I phase, which is very close 
to that of bulk MnO(111), one might suggest that these two interfacial phases are related to a MnO(111)-like structure. Conversely, the c(4$\times$2) 
phase and by extension the CHEV-I structure might be linked to MnO(100) by analogy with the c(4$\times$2)-$\rm Ni_3O_4$/Pd(100), precursor of 
NiO(100)\cite{nioleed,nioexp} and obtained under similar coverage conditions. 
To seek support for this assignment we have calculated the dipole active modes for two free standing MnO(100) and MnO(111) models containing a 
single manganese layer conveniently oxidized. The calculated phonon losses are reported in Fig.\ref{fig:3} along with a sketch of the 
corresponding phonon eigenvectors. Indeed, the resulting phonon frequencies, 31 meV and 70-75 meV, unambiguously distinguish two distinct 
vibrational modes characteristic of the (100) and (111) stacking, respectively. For these specific models PBE and HSE provide essentially 
the same description: the main MnO(100) dipole active mode at 31 meV originates from the out-of-plane opposite displacements of Mn and O atoms 
laying on the same plane, whereas the MnO(111) peak at about 75 meV arises from the antiparallel movement of adjacent Mn and O layers.
It is therefore natural to associate the calculated (100) and (111) dipole modes with the experimentally observed O-poor (43-45 meV) and O-rich (70 meV) 
phonon peaks. Consequently, with a certain degree of confidence we conclude that the square (100) and hexagonal (111) MnO thin oxide building units 
will remain relevant in the interfacial systems grown on Pd(100). Therefore they can be taken as good starting models in the search for realistic 
structures of the individual phases under consideration, namely the c(4$\times$2) and CHEV-I MnO(100)-like structures and the HEX-I and HEX-II 
MnO(111)-like phases. 

\subsection{MnO(100)-like structures: c(4$\times$2) and CHEV-I}
\label{two2}

\begin{figure*}
\begin{minipage}[b]{0.48\linewidth}
\centering
\includegraphics[clip=,width=1.0\linewidth]{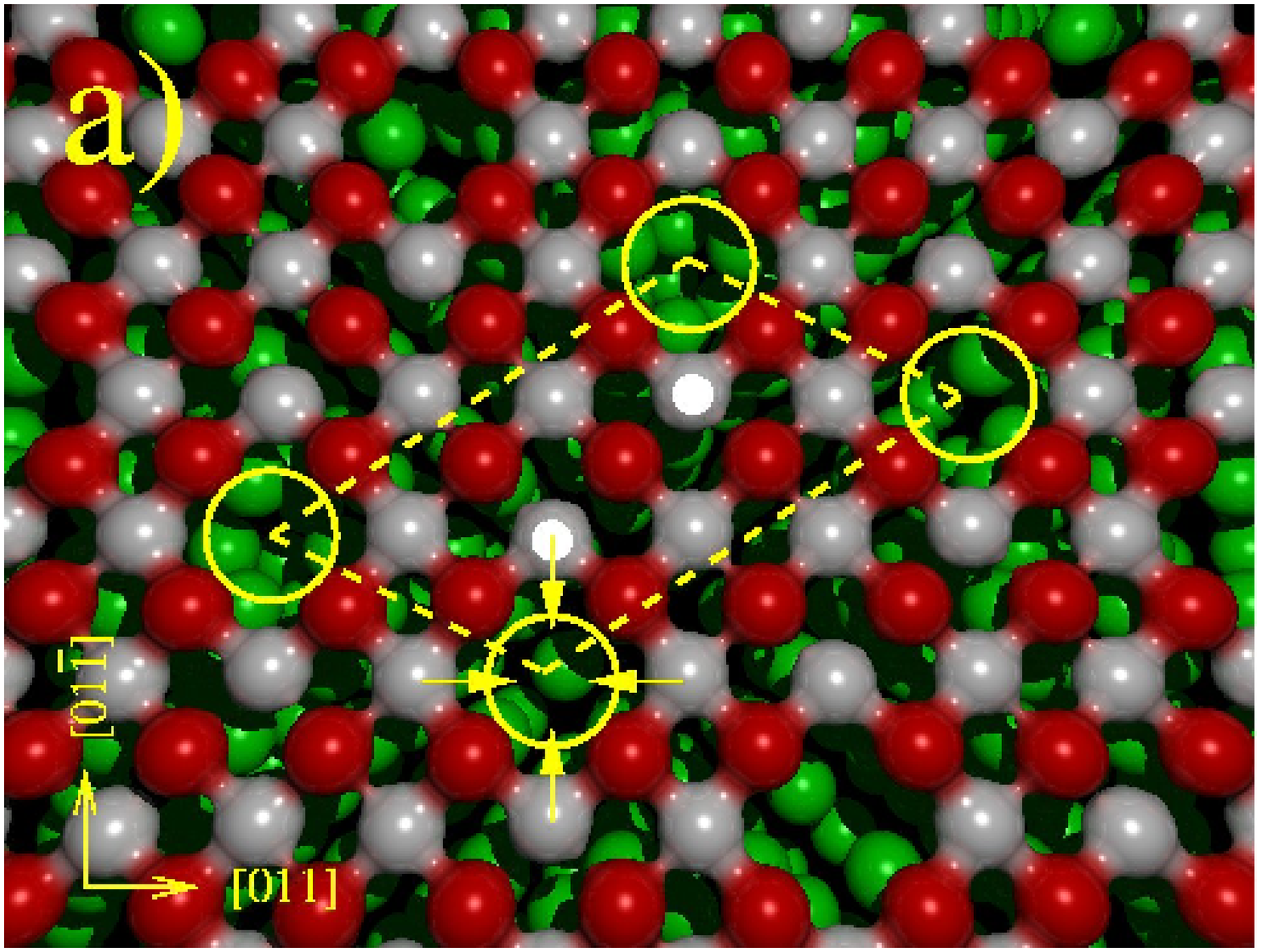}
\end{minipage}%
\hspace{0.5cm}
\begin{minipage}[b]{0.48\linewidth}
\centering
\includegraphics[clip=,width=0.75\linewidth]{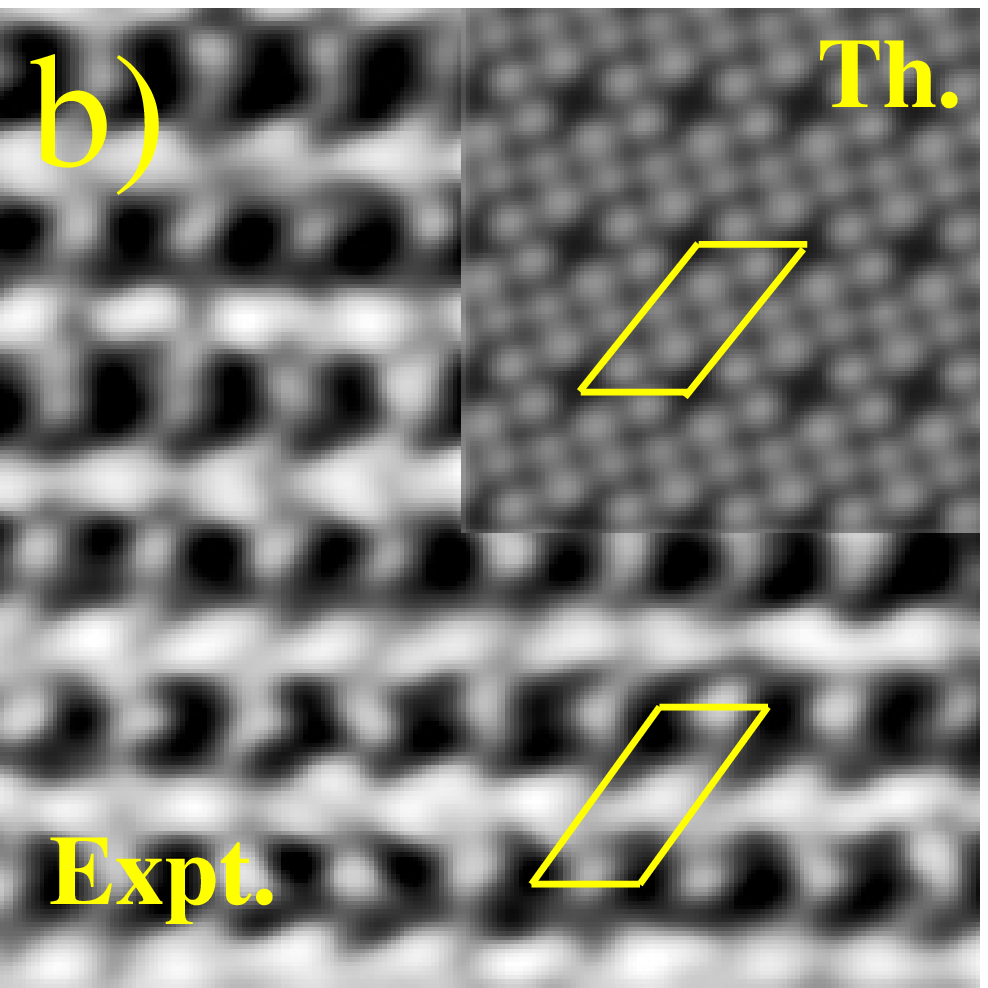} 
\end{minipage}
\begin{minipage}[b]{0.98\linewidth}
\centering
\vspace{1.0cm}
\includegraphics[clip=,width=0.6\linewidth]{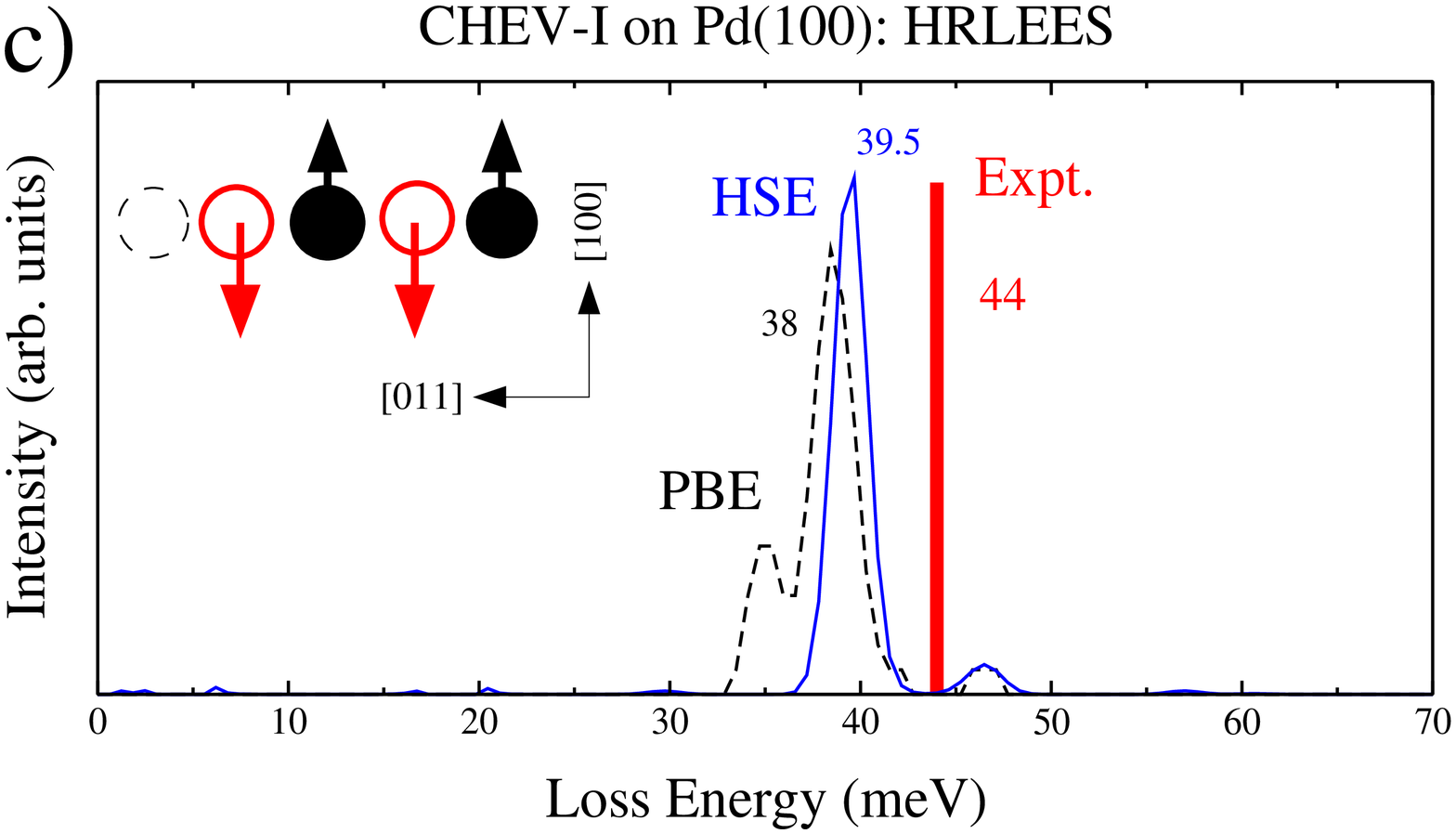}
\end{minipage}
\caption{
(a) Top view of the geometrical model for the Pd(100) supported CHEV-I $\rm Mn_6O_7$ phase: red (dark gray) spheres: O atoms;  light gray spheres: Mn atoms; 
green (gray) small spheres: Pd atoms underneath. Dashed lines delimit the 2D unit cell and circles highlight the position of the vacancies. 
(b) Experimental and simulated STM images of the CHEV-I phase; sample bias U = +1.0 V; tunneling current I =0.13 nA. The simulated STM image 
has been calculated considering tunneling into empty states between 0 and +1.0 eV. Unit cells are indicated by full lines.
(c) Comparison between the measured HREELS phonon value (vertical bar) and PBE (dashed line) and HSE (full line) predicted dipole active modes for 
the CHEV-I phase. The inset schematically depicts the atomic displacements giving rise to the calculated phonon peaks.
Mn and O atoms are sketched as filled and empty circles, respectively. The dashed line circle denotes a vacancy.}
\label{fig:6}
\end{figure*}

LEED and STM experiments have shown that an interfacial phase with c(4$\times$2) symmetry can be stabilized on Pd(100) under suitable 
conditions\cite{stmli}.  This superstructure appears to be very similar to the recently observed c(4$\times$2)-$\rm Ni_3O_4$ obtained upon reactive 
evaporation of nickel on Pd(100)\cite{nioexp, nioleed, nioth}, which has been interpreted as a compressed epitaxial NiO(100) monolayer with a 
rhombic c(4$\times$2) array of nickel vacancies. The analogy is not surprising if one considers the nature of the two oxide systems. 
Indeed, strong similarities exist between NiO and MnO in terms of structural properties 
(same rhombohedrally distorted fcc structure), magnetic ordering (same antiferromagnetic type II spin ordering) and electronic character 
(partially filled localized $d$ states). Furthermore, bulk-like MnO(100)\cite{1001} and NiO(100)\cite{spaleed} films can be both grown 
on a Pd(100) substrate.

Prompted by the strong similarities between the two c(4$\times$2) phases and guided by the structural analysis on 
the c(4$\times$2)-$\rm Ni_3O_4$/Pd(100) phase\cite{nioexp, nioleed, nioth} we have compared the relative stabilities of a number of different 
models resembling the $\rm Ni_3O_4$-c(4$\times$2) geometry. The models are constructed by placing a compressed MnO(100) monolayer on top of a 
four layer Pd(100) substrate and creating Mn vacancies with a rhombic (RH) c(4$\times$2) symmetry. The lateral compression compensates for the positive 
mismatch between MnO(100) ($a_{\rm MnO}$=3.09 \AA, PBE) and Pd(100) ($a_{\rm Pd}$=2.79 \AA, PBE). Three options have been considered concerning the 
lateral registry: (RH1) oxygen atoms placed on top of Pd with the manganese species sitting in hollow sites; (RH2) the configuration opposite to RH1 
with manganese arranged on top of Pd; and (RH3), with both Mn and O in bridge sites. As for the magnetic ordering, we have explored the FM arrangement 
and two different antiferromagnetic configurations of the Mn spins. A detailed description of the calculations will be given elsewhere\cite{c4x2}. 
For the present purposes it is essential to point out that the FM RH1 structure, schematically depicted in Fig. \ref{fig:4}(a), results as the most 
favorable one, at least 200 meV more stable than all other considered models. The fully optimized structure consists of a corrugated MnO(100) adlayer 
whose average distance from the substrate (2.42 \AA) is 22\% larger than the Pd(100) interlayer spacing. The corrugation of $\approx$ 0.23 \AA~arises 
from the outward displacement of the oxygen atoms. In the layer plane we find a negligible displacement of the oxygen atoms from their on-top positions, 
but a small shift (0.16 \AA) along the [01$\bar{1}$] direction experienced by the Mn atoms. This displacement towards the vacancy clearly is the result of 
the free space and charge left by the missing manganese and acts to release the compressive strain in the Mn-O overlayer. The rearrangements of 
the Mn species, also highlighted by the full line in Fig. \ref{fig:4}(a), results in a zig-zag Mn-Mn chain along [011] which was also observed 
in the c(4$\times$2)-$\rm Ni_3O_4$ system. 

In Fig. \ref{fig:4}(b)-(d) we compare experimental and calculated STM images of c(4$\times$2)-$\rm Mn_3O_4$.
The close resemblance between the STM simulation (inset (c)) and the high resolution STM image (Fig. \ref{fig:4}(d)) allows us to discuss 
further atomic details of the RH1 model. Atomic resolution reveals that the RH1 unit cell consists of different sub-units made up of bright spots 
with different contrast and black holes spanning the 2D unit cell. The rhombic array of dark depressions, separated by a distance of 5.58 \AA, 
reflects the network of Mn vacancies encircled in Fig. \ref{fig:4}(a), whereas the light protrusions result from the remaining Mn species.
The strong structural rearrangement induced by the lattice mismatch and by the creation of a Mn vacancy is reflected in the STM spots: 
the single bright features can be assigned to the Mn species sandwiched between two vacancies, whereas the light rows developing along the [011] 
direction are interpreted as zigzagging Mn-Mn chains embedded in a regular oxygen array (laying $\approx$ 0.23 \AA~above). 

Further support for the proposed c(4$\times$2)-$\rm Mn_3O_4$ surface structure is provided by the comparison of the calculated 
phonon spectrum with the HREELS measurements (Fig. \ref{fig:4}(e)). Remarkably, the single peak structure centered at $\approx$ 43.5 meV revealed 
by HREELS is very well reproduced by our calculations. Although standard DFT underestimates the phonon energy by $\approx$ 7 meV, the inclusion of 
a portion of Hartree-Fock exchange within the HSE method improves significantly the agreement with the experimentally measured value. The 
diagonalization of the dynamical matrix enables to assign this specific phonon mode to the anti-phase out-of-plane vibrations of surface O and 
Mn atoms, as schematically depicted in the inset of Fig. \ref{fig:4}(e). This vibrational mode is very similar to that found for the MnO(100) 
unsupported thin film (Fig. \ref {fig:3}) thus reflecting the MnO(100)-like nature of the RH c(4$\times$2)-$\rm Mn_3O_4$ phase.

\begin{figure}
\centering
\includegraphics[clip=,width=0.9\linewidth]{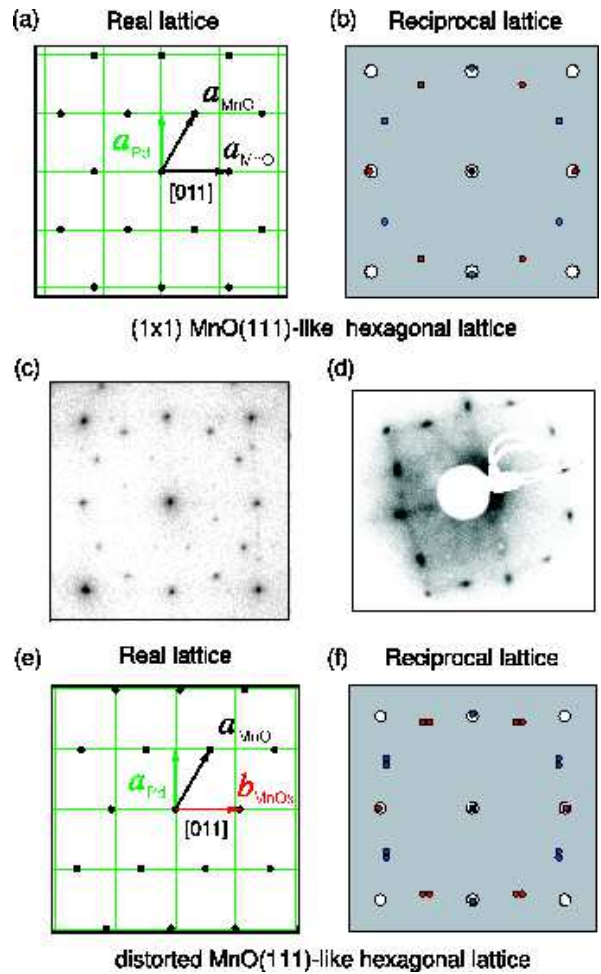}
\caption{
(a) Real and (b) reciprocal lattice for a (1$\times$1) MnO(111)-like hexagonal structure on a Pd(100) surface. 
The lattice parameter of the overlayer is assumed to be the measured bulk value $a_{\rm MnO}$= 3.14 \AA, whereas $a_{\rm Pd}$=2.75 \AA. 
In (a) only one of the two symmetry domains is reported for clarity. (c) SPA-LEED two-dimensional pattern measured at E=90 eV, for the undistorted 
HEX-I phase. (d) LEED pattern of the distorted HEX-I phase, recorded at E=96 eV. (e) Real  and (f)  reciprocal lattice associated with the distorted 
hexagonal MnO(111)-like phase. Only one symmetry domain is shown in (e) for clarity, for which the lattice parameter in the [011] direction of the 
distortion is $b_{\rm MnO_x}$=2.94 \AA. In total, four symmetry domains contribute to the reciprocal lattice in (f): two are obtained from panel (e) 
with $a_{\rm MnO}$ at either +60\ensuremath{^\circ} or -60\ensuremath{^\circ} from $b_{\rm MnO_x}$, and the other two are obtained by rotating 
$a_{\rm MnO}$ by 90\ensuremath{^\circ} relative to the substrate mesh. }
\label{fig:7}
\end{figure}

Despite the striking similarity between the two c(4$\times$2) monolayer structures formed by Ni and Mn oxides on Pd(100), precursor phases of NiO(100) 
and MnO(100), an important difference can be highlighted when considering the lattice mismatch between the two bulk oxides and the Pd(100) substrate. 
The experimental mismatch is in fact $\approx$7\% and 14\% for NiO(100) and MnO(100), respectively, indicating that the c(4$\times$2) formed by Mn oxide 
is subjected to a much higher compressive stress. This is reflected in the experimental STM images (Fig. \ref{fig:4}(b)), which show ordered regions 
bearing the c(4$\times$2)-$\rm Mn_3O_4$ periodicity separated by regions of defects. The incorporation of extensive defect areas into the film 
presumably arises from the release of the high compressive strain. The latter might also explain the tendency of the c(4$\times$2)-$\rm Mn_3O_4$ 
structure to nucleate preferentially near step edges. Moreover, we have observed that a remarkably well-ordered c(4$\times$2) overlayer forms on 
Pd(1 1 19) \cite{vic}, where (100) terraces exposing 10 atomic rows are separated by a regular array of monoatomic steps. Clearly, the steps  provide a 
suitable means of lateral relaxation and thus enable to achieve stress relief in a natural way. It is also worth noting that the high surface stress in 
the c(4$\times$2) phase may be responsible for its tendency to transform into a {\em stripe} phase (Fig. \ref{fig:1}), which can be interpreted 
as a distorted c(4$\times$2) and is obtained at comparable values of the oxygen chemical potential\cite{stmli}.

We now turn our attention to the CHEV-I structure which is adjacent to the c(4$\times$2) in the phase diagram and is obtained by lowering the $\rm O_2$ 
pressure during the preparation. This structure can be prepared as a single phase by post-oxidation method, as assessed by LEED and STM. However, the STM 
images also show that it can coexist locally with both the c(4$\times$2) and CHEV-II phases, as illustated in panel (a) of Fig. \ref{fig:5}. Here, 
it appears clearly that the c(4$\times$2) and CHEV-I structures transform smoothly into each other, a strong indication that they possess not only
similar energetics but also common structural features. Indeed, as it can be seen by inspection of the real space models in Fig. \ref{fig:5}(b) 
and (c), the CHEV-I periodicity can be naturally obtained from the c(4$\times$2) by extending one lattice vector to a neighboring  anti-phase 
position, thus yielding a structure with the matrix notation reported in Fig. \ref{fig:5}(c). As discussed above, the periodicity of the 
c(4$\times$2) model is given by the Mn vacancies. Within this picture, the CHEV-I structure may be described in terms of the propagation of a Mn 
vacancy as indicated by the red arrow in Fig. \ref{fig:5}(c). Accordingly, the stoichiometry becomes $\rm Mn_6O_7$, i.e. closer to that of MnO. 
Compared to the c(4$\times$2) phase, the CHEV-I structure exhibits a much higher degree of long range order, with large defect-free areas extending 
over typical distances of 200 \AA~and interrupted by ordered domain boundaries, which convey the characteristic chevron-like appearance \cite{stmli}(see 
also Fig. \ref{fig:5}(a)). 

Following the above arguments and encouraged by the detailed description of the c(4$\times$2) phase in terms of the 
RH1 $\rm Mn_3O_4$ model, we tested the vacancy propagation model adopting the above mentioned $\rm Mn_6O_7$ structure.
This model is constructed by removing 14\% of Mn ions from the compressed MnO(100) layer as sketched in Fig. \ref{fig:6}(a).
The geometrical optimization yields again an oxygen terminated surface with a corrugation of 0.33 \AA.
In analogy with the c(4$\times$2)-$\rm Mn_3O_4$ structure we observe a small displacement 
(0.13 \AA) of manganese atoms towards the vacancy (see arrows in Fig. \ref{fig:6}(a)).
Figure \ref{fig:6}(b) shows a comparison between the experimental and simulated STM images.
Within the limits of the experimental resolution the agreement is satisfactory. 
The dark circular depressions in the image represent the position of the Mn vacancies 
and determine the 2D unit cell of this phase, as outlined by the full lines. The bright protrusions are
clearly due to manganese atoms, suggesting that electronic effects contribute predominantly to the STM topography. 
Two distinct bright features are observed in the experimental picture, one large spot 
in the center of the unit cell and weaker flecks on the edges. The simulated STM image together with the 
optimized geometrical data allow for an atomically resolved identification of these two kinds of features. 
The manganese sublayer is itself slightly corrugated, due to a small upward shift (0.05 \AA)
of the Mn atoms laying closer to the vacancy and aligned in the [01$\bar{1}$] direction. These atoms, outlined by filled dots
in Fig. \ref{fig:6}(a), are responsible for the two large bright spots in the simulated STM image 
and should be connected to the lines of bright protrusions seen in the experimental STM image. The remaining 
spots (four per unit cell) arise from the lower manganese atoms. In the experimental STM image their intensity partially 
mixes up with the topmost Mn spots thus contributing to the large bright feature centered in the cell 
and partially providing the weaker spots along the edges. 

Finally, our DFTh calculations establish (Fig. \ref{fig:6}(c)) that the proposed $\rm Mn_6O_7$ model is characterized by 
a single dipole active mode located at around 40 meV, only 4 meV lower than the measured loss energy. This peak reflects 
the vibrational character of the parent MnO(100) not only in terms of phonon loss energy, very similar to the c(4$\times$2)-$\rm Mn_3O_4$ value, 
but also in terms of the atomic displacements producing this vibration. As sketched in the inset of Fig. \ref{fig:6}(c), the peak originates from 
the anti-phase out-of-plane movements of manganese and oxygen atoms.

\subsection{MnO(111)-like structures: HEX-I and HEX-II}
\label{two3}

In the oxygen-rich regime, the experiments show that the most stable phase in a broad range of pressures above 5$\times$10$^{-7}$ mbar is 
the HEX-I phase, which may be linked to MnO(111). The structure of bulk MnO(111) is made out of alternating O and Mn 
layers, having in-plane hexagonal arrangement with lattice constant $a_{\rm MnO}$=3.14 \AA~(see Fig. \ref{fig:7}(a)). A perfect MnO(111) film would give 
rise to a hexagonal reciprocal space pattern with a lattice constant 4$\pi$/3$a_{\rm MnO}$ = 2.30 \AA$^{-1}$, which is very close to the reciprocal Pd(100) 
lattice parameter (2$\pi$/$a_{\rm Pd}$ = 2.28 \AA$^{-1}$). The resulting LEED pattern averaged over two hexagonal domains rotated by 90\ensuremath{^\circ}, 
which account for the different symmetry of substrate and overlayer, is a circular array of 12 extra-spots superimposed on the (10) spots of the 
substrate, as illustrated in Fig. \ref{fig:7}(b). Such a LEED pattern has been indeed observed in the experiments, as demonstrated in 
Fig. \ref{fig:7}(c). Here, additional faint spots with c(6$\times$2) symmetry are also observed, which arise from the coincidence 
mesh between substrate and overlayer\cite{8x2}.

\begin{figure*}[htb] 
\begin{minipage}[b]{0.48\linewidth}
\centering
\includegraphics[clip=,width=1.0\linewidth]{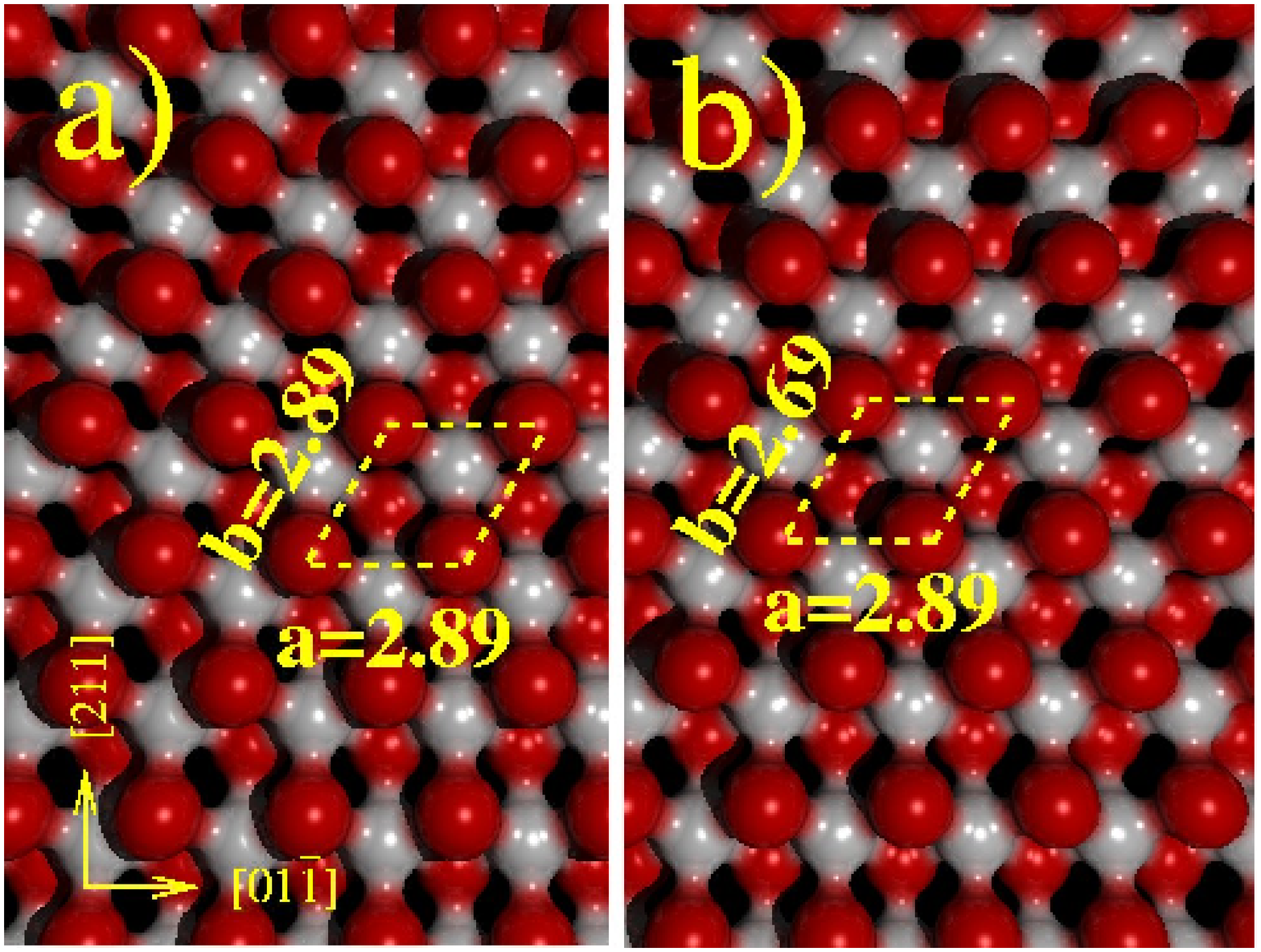}
\end{minipage}%
\hspace{0.5cm}
\begin{minipage}[b]{0.48\linewidth}
\centering
\includegraphics[clip=,width=0.75\linewidth]{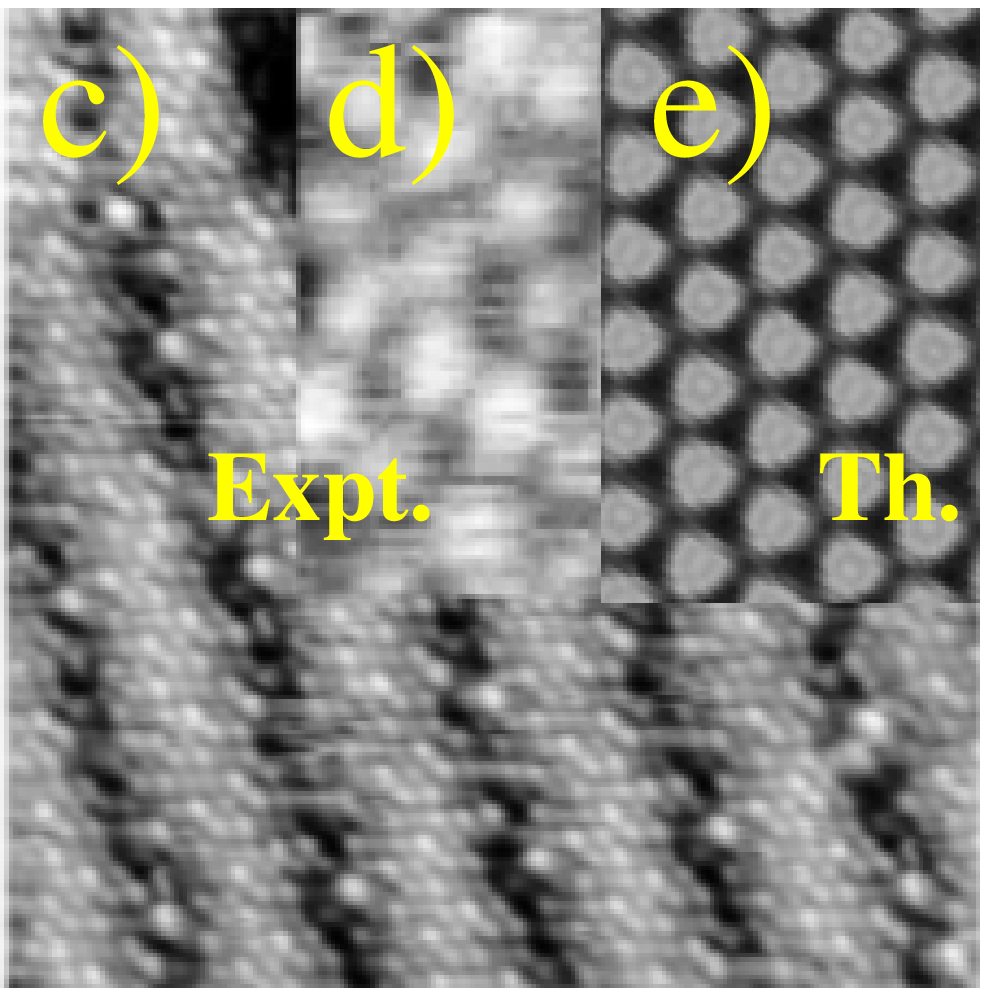}     
\end{minipage}
\begin{minipage}[b]{0.98\linewidth}
\centering
\vspace{1.0cm}
\includegraphics[clip=,width=0.6\linewidth]{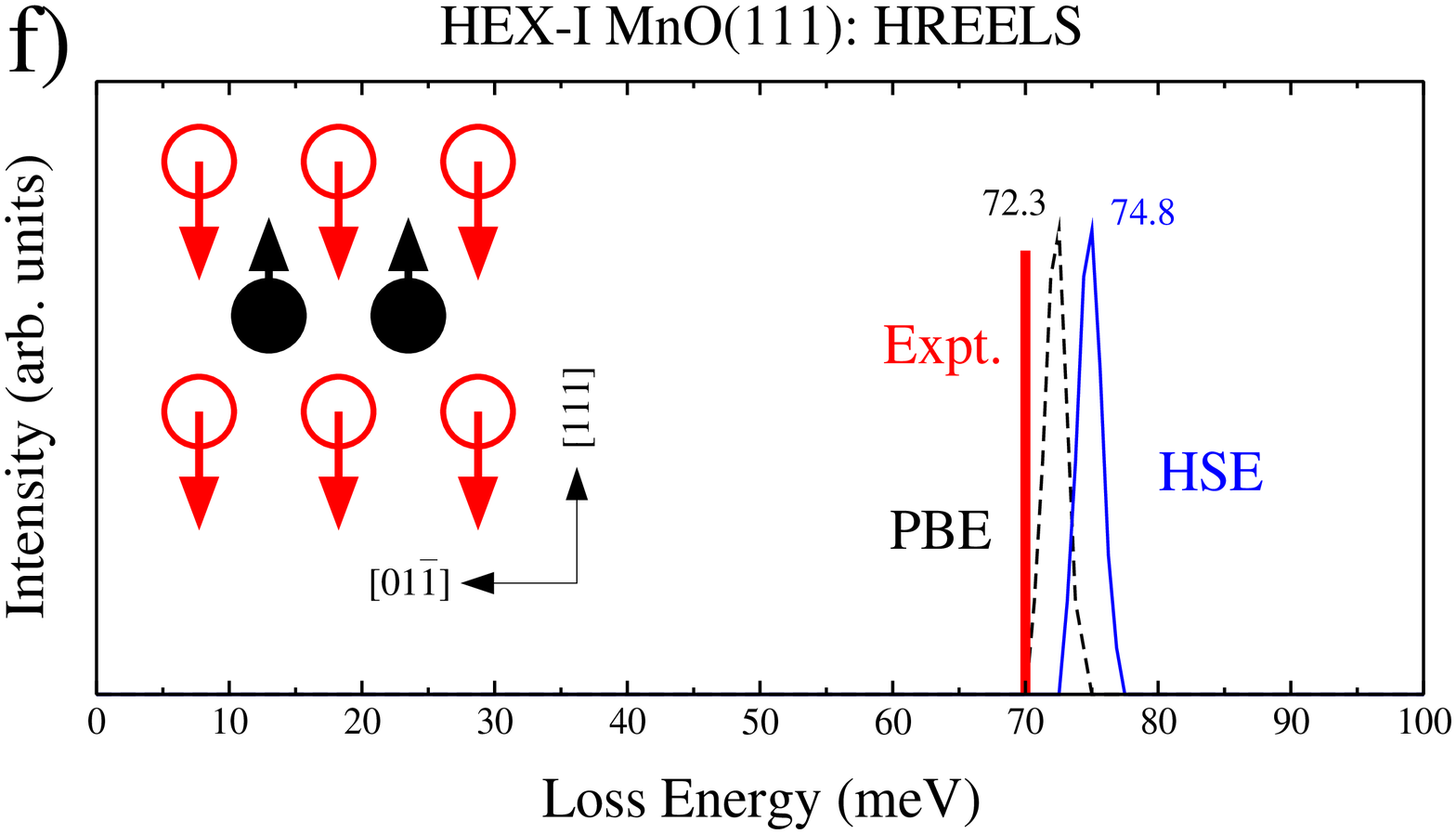}     
\end{minipage}
\caption{
(a-b) Top view of geometrical models for unsupported (a) HEX-I MnO(111) and distorted (b) HEX-I MnO(111) phases: red (dark gray) spheres: O atoms;  
light gray spheres: Mn atoms. Dashed lines delimit the 2D unit cell. (c-e) Experimental (c,d) and simulated (e) STM images of HEX-I; 
Experimental images: sample bias U = +0.5 V (c), +0.6 V (d); tunneling current I =0.13 nA (c), 0.15 nA (d). 
The simulated  STM image has been calculated considering tunneling into empty states 
between 0 and +0.5 eV. (f) Comparison between the measured HREELS phonon value (vertical bar) and PBE (dashed line) and HSE (full line) predicted 
dipole active modes for the HEX-I phase. The inset schematically depicts the atomic displacements giving rise to the calculated phonon peaks.
Mn and O atoms are sketched as filled and empty circles, respectively.}
\label{fig:8}
\end{figure*}

A similar LEED pattern has been also obtained at comparable coverage for oxidized Mn on Rh(100) and attributed to MnO(111) \cite{Rhjap}. For 
Mn oxide nanolayers on Pd(100), however, it has not been observed always reproducibly in the different experimental systems. It appears that 
this structure is favored at a slightly lower coverage (0.5-0.6 ML) or after repeated oxidation cycles. Thus, the experimental evidence suggests 
that this "perfect" (1$\times$1) structure may have limited stability on Pd(100) or may be kinetically stabilized. In contrast, the LEED pattern 
shown in Fig. \ref{fig:7}(d) was reproducibly obtained after oxidation at high pressure of 0.75 ML. Here, the characteristic elongation of the 
overlayer spots indicates a distortion of the "perfect" (1$\times$1) structure. It consists of either a contraction of one lattice vector or 
a reduction of the 60\ensuremath{^\circ} angle of the 2D hexagonal cell. The first type of distortion explains successfully the Moir\'e pattern 
observed in the STM images (see Fig. \ref{fig:8}(c)): by reducing the lattice vector along [011] to $b_{\rm MnO_x}$ = 2.94~\AA (Fig. \ref{fig:7}(e))
the experimentally observed Moir\'e periodicity of $\approx$~22~\AA~along this direction is correctly reproduced\cite{stmli} and the LEED spots appear 
elongated or split (Fig. \ref{fig:7}(f)). By inspection of  Fig. \ref{fig:7}(a) and (e), one may infer that the closer matching between the 
overlayer and substrate meshes is the driving force for the distortion: the initial mismatch of 14\% along the [011] direction is in fact reduced 
to ($b_{\rm MnO_x}-a_{\rm Pd}$)/$a_{\rm Pd}$=7\%. This better epitaxial relationship is presumably at the origin of the increased stability of the distorted 
HEX-I phase. 

\begin{figure}
\centering
\includegraphics[clip=,width=1.0\linewidth]{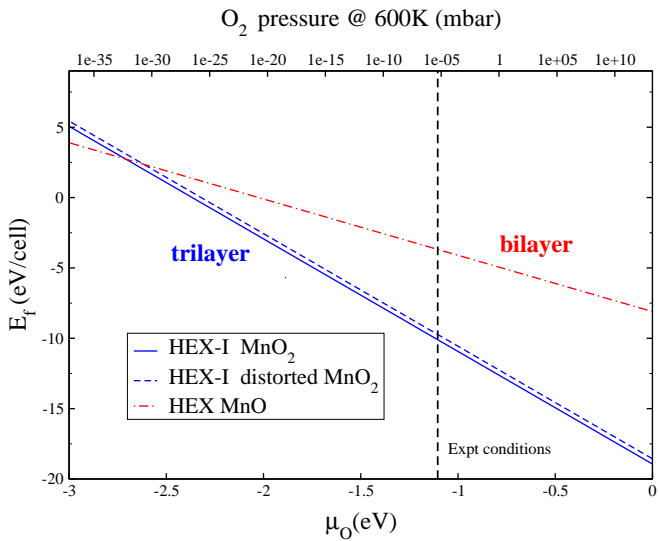}
\caption{Thermodynamic DFT phase stability diagram of explored HEX-I models in equilibrium with an O particle reservoir controlling the chemical
potential $\mu_O$. The top scale is converted into oxygen pressure.
}
\label{fig:9}
\end{figure}

As already pointed out in Section \ref{one2} the computational modeling of the HEX-I and HEX-II phases has to be done with the unsupported setup, 
namely considering only free-standing thin layers. The most natural way to 
construct such hexagonal thin Mn-O layers is by cleaving MnO perpendicular to the [111] direction. In the rocksalt (111) stacking 
the smallest blocks containing 1 ML of Mn atoms are either a Mn-O bilayer or an O-Mn-O trilayer. We have first studied the relative stability of 
these two configurations by minimizing the formation energy with respect to the 2D hexagonal lattice constant. The computed energies E$_f$, plotted 
in Fig. \ref{fig:9}, clearly show that the trilayer structure, with a formal stoichiometry of $\rm MnO_2$, is by far more favorable in a wide 
portion of the phase diagram than the bilayer MnO structure. Only in the oxygen-poor regime the bilayer gains stability with respect to the trilayer. 
However, the latter conditions are unrealistic compared to the experimental situation ($\approx 10^{-6}$ mbar). 

A rigid sphere sketch of the $\rm MnO_2$ trilayer is given in Fig. \ref{fig:8} (a). As a direct consequence of the free standing nature 
the minimized planar lattice constant, $a=2.89$ \AA, is considerably smaller than the corresponding calculated bulk lattice constant of 3.09 \AA. 
For the same reason, the optimized interlayer distance, 0.95 \AA, is found to be 25 \% shorter than the bulk value. In a second step, in order to 
simulate the experimentally observed distorted hexagonal structure we have computed the energy required to distort the perfect hexagonal trilayer 
by shrinking the lattice constant $b$ by 7\%. We found that the distorted trilayer is only 70 meV less stable than the perfect hexagonal trilayer, 
as shown in the phase diagram of Fig. \ref{fig:9}. Therefore, although we find that the perfect hexagonal structure is the most favorable, 
only a low energy cost is required to convert it into a distorted hexagonal structure. Considering that our model neglects the effect of the 
Pd(100) substrate and that the distortion allows for a better lattice matching with the substrate, we conclude that the distorted trilayer model
of the HEX-I is also supported by our calculations. We have tested the soundness of the proposed $\rm MnO_2$ trilayer model by comparing the 
experimental STM and HREELS data with the theory. The results are presented in Fig. \ref{fig:8} (c-f). As evident from Fig. \ref{fig:8} (c) 
the HEX-I phase displays the Moir\`e-like superstructure already discussed \cite{stmli}. The atomically resolved STM shown in panel (d) provides 
information on the local hexagonal symmetry of the array of bright protrusions, while panel (e) gives the result of the simulation. We find a 
fair agreement between the two images, and the theoretical data allow us to assign the bright spots to the topmost oxygen atoms. Convincing additional 
support for our model comes from the comparison between the experimental and calculated phonon spectra of Fig. \ref{fig:8} (f). DFT and DFTh 
locate the distinct dipole active mode between 72 and 75 meV, which is only a few meV higher than the measured value of 70.5 meV. The displacements 
corresponding to this mode, sketched in the inset, clearly show that this particular vibration is linked to the alternately stacked hexagonal layers of 
manganese and oxygen atoms along the direction perpendicular to the surface and is not compatible with the MnO(100)-like monolayer.
It is also worth noting that the trilayer model is consistent with the O $1s$ core level photoemission data presented in Ref. \cite{stmli},
which hint at the presence of two O $1s$ components separated by rougly 0.4 eV in binding energy. In fact, the stronger photoemission component at 529.1 eV
can be assigned to the surface O layer, whereas the weaker line at higher binding energy can be related to the O layer at the interface
with the Pd(100) substrate, in close analogy with previous assignments for trilayer oxide systems\cite{rh1, rh2}. 

\begin{figure}[htb]
\centering
\includegraphics[clip=,width=1.0\linewidth]{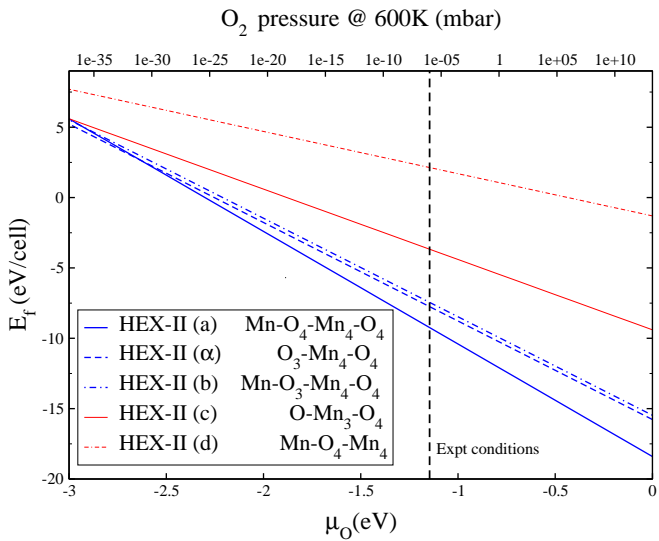}     
\caption{
Thermodynamic DFT phase stability diagram of explored HEX-II models in equilibrium with an O particle reservoir controlling the chemical potential 
$\mu_O$. The top scale is converted into oxygen pressure.  }
\label{fig:10}
\end{figure}

The O-Mn-O MnO(111) trilayer model is the natural starting point for the structural understanding of the HEX-II phase which is obtained from the 
HEX-I phase upon reduction of the oxygen partial pressure. We rely on our previous work on the reconstruction of the bulk-terminated MnO(111) 
surface\cite{111}, which allows us to restrict significantly the structural search. The STM and LEED analysis provide clear evidence for a well 
ordered 2D hexagonal unit cell with lattice constant $a\approx 6.0$ \AA~(Ref. \cite{stmli}), corresponding to a compressed (2$\times$2) 
reconstruction of the MnO(111) termination. Our DFT calculations revealed that all reconstructions with (2$\times$2) periodicity are more stable 
than others with (2$\times$1) or ($\sqrt{3}\times\sqrt{3}$) symmetry. Amongst the explored (2$\times$2) structures we found that an oxygen terminated 
octopolar reconstruction and a so-called $\alpha$ phase display the lowest surface energies in a wide range of oxygen partial pressure. Hence, we 
have focused our search on related (2$\times$2) structures, limiting the exploration to thin layers containing $\lesssim$ 1ML Mn.  
The octopolar reconstruction contains local pyramidal structures obtained by removing one quarter of atoms from the sub-surface and three 
quarters from the surface layer. It consists of triangular based pyramids with either $\rm O-Mn_3-O_4$ or $\rm Mn-O_3-Mn_4$ stacking, respectively. 
Instead, the $\alpha$ structure is obtained by removing the topmost atom from the octopolar reconstruction, resulting in a $\rm Mn_3-O_4$ (Mn-terminated) 
or $\rm O_3-Mn_4$ (O-terminated) stoichiometry. In addition to the stoichiometrically perfect octopolar and $\alpha$ reconstructions we have also 
considered slightly modified models, obtained by suitable addition or removal of one surface or sub-surface atom. In Fig. \ref{fig:10} we 
summarize the phase diagram of the most stable models explored. We first point out that all quasi-bilayer models (with ad-atoms) such as HEX-II (c) 
and HEX-II (d) are less favorable. Conversely, the models that maintain a trilayer nature compatible with the octopolar and $\alpha$ reconstructions 
are clearly preferred, such as HEX-II (a), HEX-II $\alpha$ and HEX-II (b). 

\begin{figure*}[htb]
\begin{minipage}[b]{0.48\linewidth}
\centering
\includegraphics[clip=,width=1.0\linewidth]{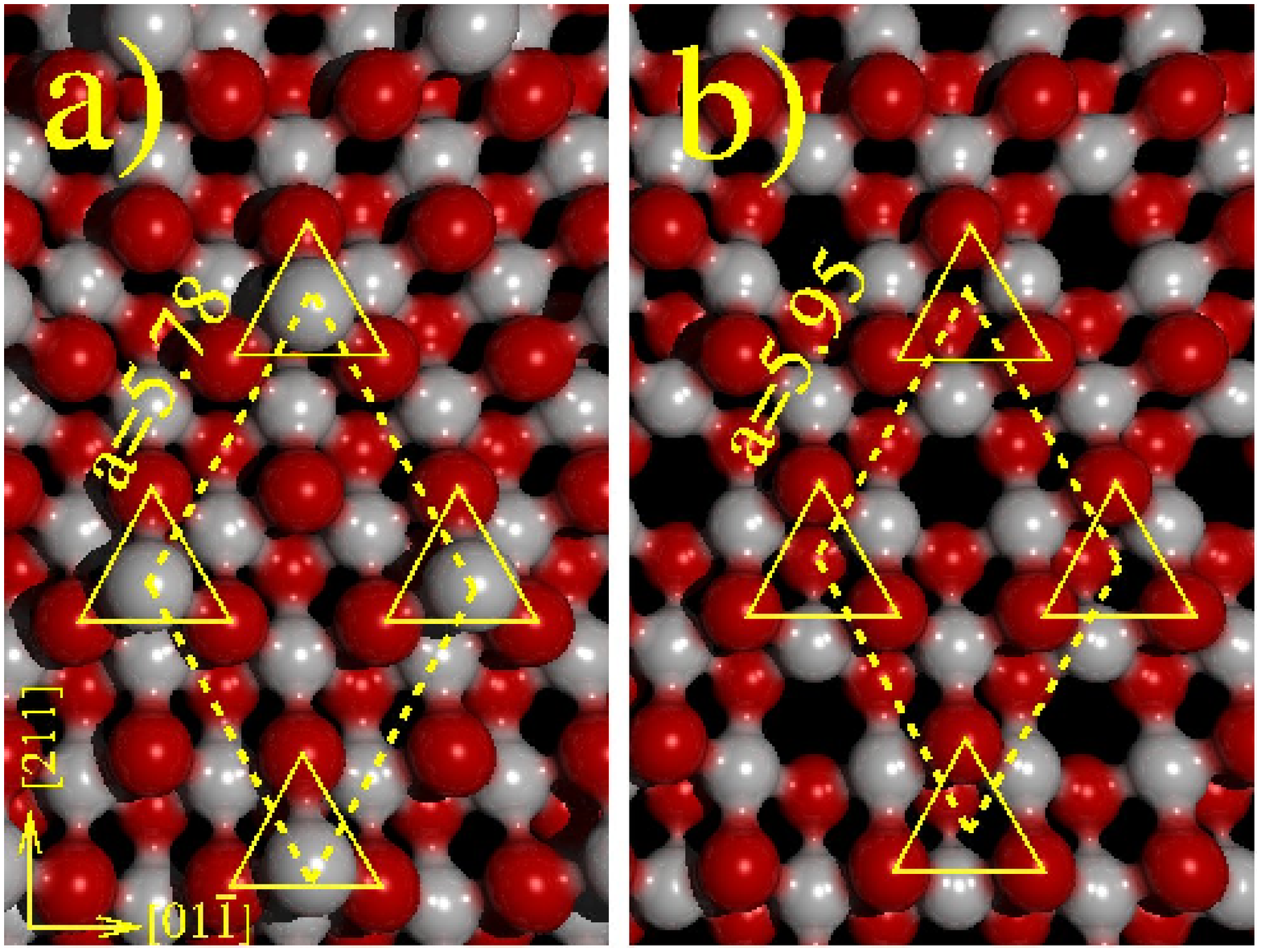}
\end{minipage}%
\hspace{0.5cm}
\begin{minipage}[b]{0.48\linewidth}
\centering
\includegraphics[clip=,width=0.76\linewidth]{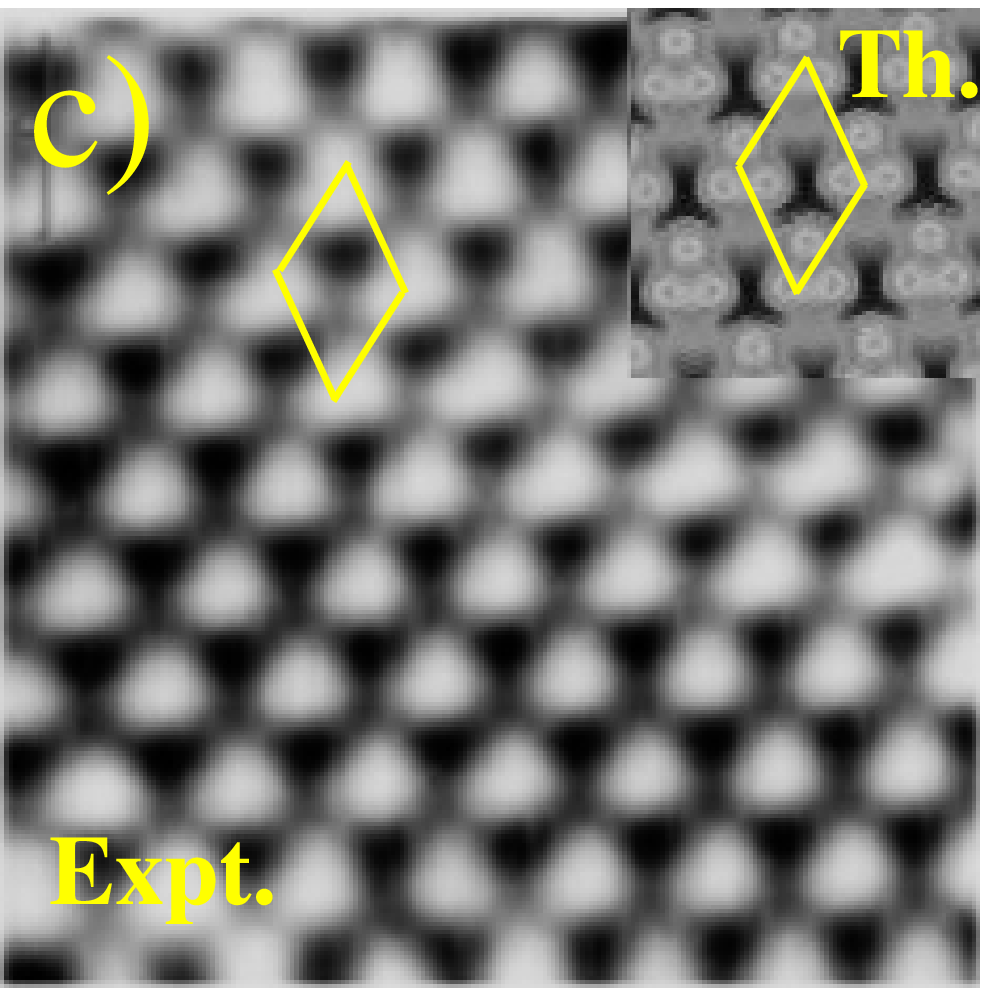}    
\end{minipage}
\begin{minipage}[b]{0.98\linewidth}
\centering
\vspace{1.0cm}
\includegraphics[clip=,width=0.6\linewidth]{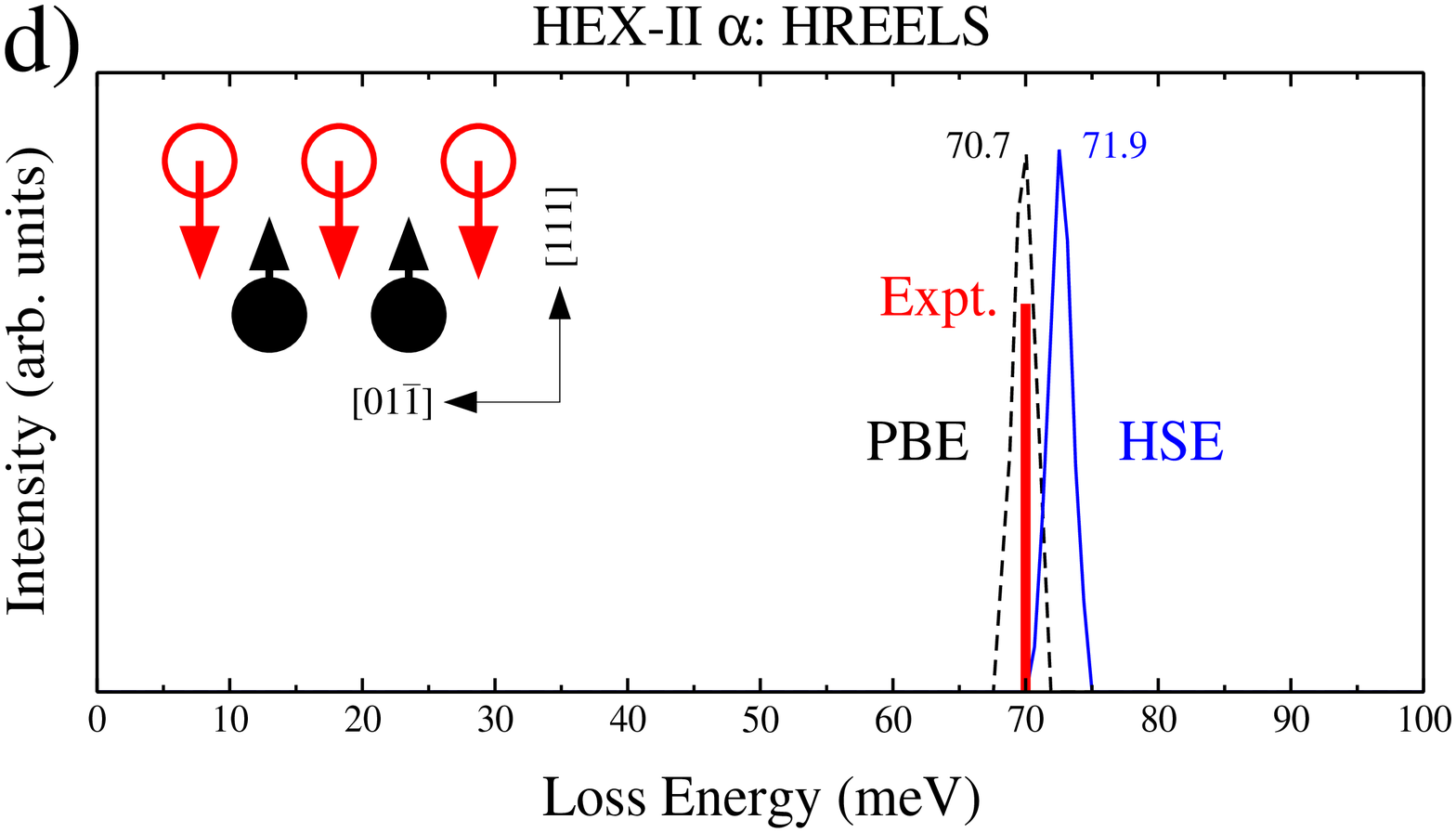}     
\end{minipage}
\caption{
(a-b) Top view of the two most stable models for unsupported HEX-II, namely (a) octopolar-like $\rm Mn_5O_8$ and (b) $\rm Mn_4O_7$-$\alpha$ phases:
red (dark gray) spheres: O atoms;  light gray spheres: Mn atoms. Dashed lines delimit the 2D unit cell, whereas triangles highlight the octopolar Mn-O$_3$ 
pyramids and the O$_3$ trimers of the $\alpha$ termination.(c) Experimental and simulated STM images of the hexagonal $\alpha$ HEX-II phase;
sample bias U = +0.75 V; tunneling current I =0.1 nA. The simulated  STM images have been calculated considering tunneling into empty states
between 0 and +0.75 eV. Unit cells are indicated by full lines. (d) Comparison between the measured HREELS phonon value (vertical bar) and
PBE (dashed line) and HSE (full line) predicted dipole active modes for the $\alpha$ HEX-II phase. The inset schematically depicts the atomic 
displacements giving rise to the calculated phonon peaks. Mn and O atoms are sketched as filled and empty circles, respectively. }
\label{fig:11}
\end{figure*}

Let us now focus on the two most stable models, HEX-II (a) and HEX-II $\alpha$, whose schematic view is given in Fig. \ref{fig:11} (a) and (b). 
HEX-II (a) is obtained by adding one manganese atom on top of the HEX-I trilayer, whereas HEX-II $\alpha$ is the stoichiometrically perfect oxygen 
terminated $\alpha$ reconstruction, built by removing one surface oxygen atom per (2$\times$2) unit cell from the HEX-I structure. At first glance 
HEX-II (a) appears to be the most favorable model, but a deeper inspection leads to a different conclusion. We first note that the Mn coverage of 
HEX-II (a) is 1.25 ML, i.e. larger than the experimental one, and that the optimized planar lattice constant for the HEX-II (a) model, 5.78 \AA, 
is smaller than the measured value of 6.0 \AA. Conversely, the O-terminated $\alpha$ model has the right Mn coverage of 1 ML, and its relaxed
lattice constant is 5.95 \AA, i.e. almost identical to the experimental value. Moreover, compelling evidence for the O-terminated $\alpha$ phase 
comes from the comparison between the simulated STM and HREELS data and the corresponding experimental results given in Fig. \ref{fig:11} (c) and 
(d), respectively. The STM image reveals a hexagonal pattern of large bright triangular intensity maxima well reproduced by the simulated STM image. 
The latter display triangular protrusions, which are ascribed to the oxygen trimers outlined in the sketch of panel (b). Instead, the simulated 
HEX-II (a) STM image (not shown) is characterized by circular bright protrusions. As for the HREELS spectrum, in analogy with its precursor, the 
HEX-I phase, the HEX-II $\alpha$ displays a single peak centered at 69.5 meV, characteristic of the MnO(111)-like structures. The simulation reproduces 
well the position of the vibrational peak and ascribes it to the concerted movement of the surface oxygen trimers against the manganese atoms underneath. 
In contrast, the calculated dipole active mode for the HEX-II (a) model is significantly higher in energy, 83 meV, and comes from a different kind of displacements, 
namely the vertical vibration of the octopolar Mn apex against the triangular oxygen basis. Taken together, it appears that the $\alpha$ structure is the 
best of the explored models in reproducing the experimental scenario in terms of structural properties, vibrational spectra and STM images. 
Although the HEX-II (a) phase is energetically competitive and even somewhat preferred, other calculated properties such as the dipole active mode 
and STM image do not match as well with the experimental data. 

\section{Summary and Conclusions}
\label{three}
The $\rm Mn_xO_y$/Pd(100) system possesses a complex and intriguing phase diagram.
In previous work\cite{1001,noimn3o4} we have described the high coverage regime (15-20 ML),
where well-ordered bulk-type MnO(100) films can be grown in a wide range of temperature and pressure, which in turn can be
converted either into MnO(111) or $\rm Mn_3O_4(001)$ structures by tuning the appropriate conditions of temperature and pressure. 
At (sub)monolayer coverages an even richer phase diagram is found, which comprises nine different structures (Fig. \ref{fig:1}). 
Here, by means of combined experimental and theoretical tools we have carried out 
a detailed investigation on four prototypical phases (c(4$\times$2), CHEV-I, HEX-I and HEX-II), which belong to two distinct oxygen 
pressure regimes and are characterized by well-defined structural and vibrational properties. 

At oxygen-poor conditions a two-dimensional c(4$\times$2)-$\rm Mn_3O_4$ structure is stable, which is described as a MnO(100) 
monolayer with a rhombic array of manganese vacancies, in close analogy to the c(4$\times$2)-$\rm Ni_3O_4$/Pd(100) 
structure \cite{nioleed,nioexp,nioth}. Upon reduction, vacancy propagation leads to the $\rm Mn_6O_7$ CHEV-I phase, also representing 
a monolayer structure. The close structural relation between CHEV-I and c(4$\times$2) is reflected in the common vibrational fingerprint, 
with a single HREELS loss peak at $\approx 44$ meV due to the anti-phase out-of-plane movements of almost co-planar Mn and O ions. 

In the oxygen-rich regime the scenario changes substantially. The observed HEX-I and HEX-II phases possess
a MnO(111)-derived O-Mn-O trilayer structure and display the same dipole active mode at $\approx$ 70 meV.
The transition between HEX-I and HEX-II is driven by the formation of oxygen vacancies inducing a so-called $\alpha$
reconstruction and changing the stoichiometry from $\rm MnO_2$ to $\rm Mn_4O_7$. 
Incidentally, the occurrence of similar trilayer models have been also invoked to describe other transition metal (Rh and Ir) 
surface oxides\cite{rh1, rh2, ir}. 
 
The correct assignment and atomic-level description of all structures is validated by the excellent agreement between experiment 
and theory in terms of structural and vibrational properties. 

\section{Acknowledgments}
This work has been supported by the Austrian Science Funds FWF within the Joint Research
Program S90 and the Science College W4, and by the 6$^{th}$ Framework Programme of the European
Community (GSOMEN).


\begin{thebibliography}{100}
\bibitem{mark1} Reported pressures were not corrected for the ion gauge sensitivity.
\bibitem{lund} We found further evidence for the existence of two different regimes in X-ray photoemission and absorption experiments with
synchrotron radiation.
\bibitem{8x2} A c(6$\times$2) coincidence net is only obtained when $a_{MnO}$ is expanded by $\approx$ 3\%. A perfect MnO(111) structure would rather
result in a c(8$\times$2) coincidence lattice.
\end{thebibliography}
\end{document}